\documentclass[]{aastex631}
\usepackage{fixltx2e}

\newcommand{\R}{\textcolor{red}}
\newcommand{\B}{\textcolor{blue}}
\shorttitle{Phaethon WISPR Update}
\shortauthors{Battams et al.}
\graphicspath{{./}{figures/}}

\begin{document}

\title{Continued PSP/WISPR Observations of a Phaethon-related Dust Trail }

\correspondingauthor{Karl Battams}
\email{karl.battams@nrl.navy.mil}

\author[0000-0002-8692-6925]{Karl Battams}
\affil{US Naval Research Laboratory, 4555 Overlook Avenue SW, Washington, DC 20375, USA}

\author[0000-0002-6420-8404]{Angel J. Gutarra-Leon}
\affiliation{George Mason University, Fairfax, VA, USA}

\author[0000-0002-8353-5865]{Brendan M. Gallagher}
\affil{US Naval Research Laboratory, 4555 Overlook Avenue SW, Washington, DC 20375, USA}

\author[0000-0003-2781-6897]{Matthew M. Knight}
\affiliation{United States Naval Academy, 572C Holloway Rd, Annapolis, MD 21402, USA}

\author[0000-0001-8480-947X]{Guillermo Stenborg}
\affil{Johns Hopkins University Applied Physics Laboratory, Laurel, MD, USA}

\author{Sarah Tanner}
\affiliation{Dundalk Institute of Technology, Dundalk, Ireland, UK}

\author[0000-0000-0000-0000]{Mark G. Linton}
\affil{US Naval Research Laboratory, 4555 Overlook Avenue SW, Washington, DC 20375, USA}

\author[0000-0003-2685-9801]{Jamey R. Szalay}
\affil{Princeton University, Princeton, NJ 08544, USA}

\author[0000-0002-6702-7676]{Michael S.P. Kelley}
\affil{University of Maryland, College Park, MD 20742, USA}

\author[0000-0001-9027-8249]{Russell A. Howard}
\affil{Johns Hopkins University Applied Physics Laboratory, Laurel, MD, USA}

\begin{abstract}

We present an update to the first white-light detections of a dust trail observed closely following the orbit of asteroid (3200) Phaethon, as seen by the Wide-field Imager for Parker Solar Probe (WISPR) instrument on the NASA Parker Solar Probe (\textit{PSP}) mission. Here we provide a summary and analysis of observations of the dust trail over nine separate mission encounters between October 2018 and August 2021 that saw the spacecraft approach to within 0.0277 au of the orbit of Phaethon. We find the photometric and estimated dust mass properties to be inline with those in the initial publication, with a visual (V) magnitude of V$\sim$16.1$\pm$0.3 per pixel, corresponding to a surface brightness of 26.1~mag~arcsec$^{-2}$, and an estimated mass of dust within the range $10^{10}$~kg -- $10^{12}$~kg depending on the assumed dust properties. However, the key finding of this survey is the discovery that the dust trail does not perfectly follow the orbit of Phaethon, with a clear separation noted between them that increases as a function of true anomaly, though the trail may differ from Phaethon's orbit by as little as 1$^{\circ}$ in periapsis.

\end{abstract}


\section{Introduction} \label{sec:intro}

The Wide-Field Imager for Parker Solar Probe \citep[WISPR;][]{Vourlidas15} instrument, operating aboard the NASA \textit{Parker Solar Probe} satellite \citep[PSP;][]{Fox15} since its launch in 2018, is designed to observe visible-light solar outflow and structures in the extended solar corona, and after just three years of operations has already returned a number of valuable results in this field \citep[e.g.][]{Hess2020,Poirier2020,Wood2020}. By virtue of its sensitive cameras and unique near-Sun, inner-solar system viewpoint, the instrument has also proved to be highly successful at imaging faint inner solar system dust structures, including proving the existence of a long-theorized dust-depletion zone near the Sun \citep{Howard2019,Stenborg2021DDZ} and providing white-light observations of a circumsolar dust ring in the orbit of Venus \citep{Stenborg2021VenusRing}. As detailed in our initial publication \citep{Battams2020} (hereafter referred to as ``Paper I''), \textit{PSP}/WISPR has also provided the first white-light observations of a dust trail appearing to follow the orbit of so-called ``activated'' asteroid 3200 Phaethon (hereafter referred to as ``Phaethon''), and presumably representing some portion of the Geminid stream.

Since its discovery in 1983, Phaethon has remained a well-studied yet poorly understood object, known to exhibit both cometary and asteroidal characteristics, and widely presumed to be parent to the Earth-crossing Geminid meteor shower \citep{IAUC3878,Gustafson1989,Williams1993}. Significant efforts have been made to study the physical properties of the asteroid such as size \citep[6~km in diameter,][]{taylor19}, spectral shape \citep[B-type, e.g.,][]{Licandro2007}, and albedo \citep[0.11,][]{Green1985}, as well as the Geminids \citep[e.g.][]{Hughes1989,Jenniskens1994,Arlt2006,Blaauw2017}, their relationship to one-another \citep[e.g.][]{Gustafson1989,Deleon2010}, and the proposed relationship to asteroids 2005 UD and 1999 YC \citep[e.g.,][]{Ohtsuka2006,Ohtsuka2008,Kareta2021}. Several Earth-based studies of Phaethon have sought to investigate the perihelion activity of the asteroid, attempting -- but failing -- to detect small fragments \citep{Jewitt2018,Ye2018} or nearby dust grains \citep{Jewitt2019} released during its close (0.14 au) passage by the Sun. Despite this, spacecraft observations of Phaethon's activity at perihelion obtained by the Sun-Earth Connection Coronal and Heliospheric Investigation \citep[SECCHI;][]{Howard2008} on NASA's {\it Solar Terrestrial Relations Observatory} \citep[{\it STEREO};][]{Kaiser07} during the asteroid's perihelion passages in 2009, 2012, 2016, and 2022, have shown that the asteroid is clearly active at perihelion, displaying a short tail and reported to be releasing mass on the order $\sim~2.5{\times}10^{8}a_1$~kg (where $a_1$ is the grain radius in mm, and noting that the size of dust released at perihelion remains unknown) \citep{Jewitt2010,Jewitt2013,Hui2017}. This is insufficient by orders of magnitude to support the best estimates for the mass of the Geminids. Shortly before submission of this manuscript, Phaethon was also observed by the Solar and Heliospheric Observatory (\textit{SOHO}) Large Angle Spectrometric Coronagraph (LASCO, \cite{Brueckner1995}) C2 coronagraph during its 2022 May 15 perihelion passage, constituting the first known detection of Phaethon by this instrument.

It has been proposed that much of the activity in certain near-Sun bodies, in particular Phaethon, may be driven by sodium emission in the absence of more volatile substances (e.g. water) \citep{Masiero2021}, a result supported by the apparent lack of sodium observed in Geminids \citep{Abe2020, Kasuga2005}. An alternative mass loss mechanism, discussed recently by \cite{Kimura2022} presents the possibility that micrometer-sized dust particles, aided by high-temperature alkali ions, may be electrostatically lofted from the surface of Phaethon around its perihelion. This mechanism was also briefly noted by \cite{Szalay2019} as potential small contributor to the near-Phaethon dust environment, though the focus of this study was on an alternate mechanism for Phaethon's mass loss - namely, micrometeorite impacts and their resulting dust ejecta - with the study finding in part that the electrostatic lofting mechanism is unlikely to provide substantial contributions to the Phaethon-Geminid dust system. Recent work by \cite{Ye2019} suggests that thermally-driven destruction of near-Sun asteroids may be a major contributor to Sun-approaching meteoroid streams, implying that such processes are not limited to cometary bodies, although the exact details of the processes leading to the ultimate destruction of these bodies remains somewhat unclear.

The Geminid meteor shower, whose nominal orbit differs from that of Phaethon, likewise is a well-studied structure, with Earth's annual intersection with the stream providing us a regular opportunity to sample a small portion. Yet our understanding of the formation and structure of the Geminids also remains poor, despite modeling efforts over several decades \citep[e.g.,][and references therein]{Jones1986}, with the stream's age generally accepted to be on the order 2~kyr, dating back to when Phaethon's perihelion was likely at its lowest value \citep{Hanus2016,MacLennan2021}. Recent optical ground-based attempts at observing the stream, or dust trails along or near Phaethon's orbit, remain unsuccessful \citep{Urakawa2002,Ishiguro2009,Hui2017}, with the only previous positive detection of such a trail coming from a re-examination of infrared observations from the {\it Cosmic Background Explorer's (COBE)} Diffuse Infrared Background Experiment (DIRBE), which revealed a faint dust trail following Phaethon's orbit \citep{Arendt2014}. While the true structure and extent of the stream remains unknown, lunar impacts of Geminids have implied that the stream may at the very least be bifurcated \citep{Szalay2018}, and beta meteoroids ejected from the Geminids have been proposed as a significant population of dust known to impact the \textit{PSP} satellite at certain portions of its orbit \citep{Szalay2021}.

In Paper I, we presented the first detection of a visible-light dust trail following the orbit of Phaethon \citep{Battams2020}, as observed by the Inner detector of \textit{PSP}/WISPR (WISPR-I) when the spacecraft approached to within 0.0765 astronomical units (au) from Phaethon's orbit during \textit{PSP}'s first solar ``encounter''. Using these observations we found the dust trail to have a visual ($V$) magnitude of 15.8~$\pm$~0.3 per pixel and a corresponding surface brightness of 25.0 mag arcsec$^{-2}$. This led to a derived estimate for the total mass of the stream to be $\sim(0.4-1.3){\times}10^{12}$~kg, which is plausibly consistent with the very low end of the assumed mass of the Geminid stream, but significantly more than believed to be produced by activity of Phaethon at perihelion. We also demonstrated that, at geocentric distances appropriate for ground-based observations, the surface brightness of this dust trail would be on the order of 31 mag arcsec$^{-2}$, which is at or beyond the limits of most terrestrial observatories and explains the lack of detections to date.

In the subsequent years \textit{PSP} has performed ten additional solar encounters including three Venus fly-bys to place the spacecraft's orbits increasingly close to the Sun, with the most recent encounter at time of submission (Encounter 11, 2022 Feb) seeing the spacecraft approach to within 0.074 au of the Sun. In each of these encounters the spacecraft has also approached closer to Phaethon's orbit, most recently to within 0.0277 au, with the dust trail described in our initial publication observed with increasing clarity in both of the WISPR cameras, but particularly the outer (WISPR-O) camera. Here we provide an update to our initial study of the trail, providing details of the photometric and morphological properties observed following the evolution of the \textit{PSP} orbit, and identifying crucial details about the orbit of the dust trail.

\section{Data and Observations} \label{sec:data}

\subsection{PSP/WISPR Overview}
The \textit{PSP} spacecraft operates on a highly elliptical, evolving orbit which sees the spacecraft gradually increase its proximity to the Sun from an initial close-encounter distance of approximately 24-million kilometers (0.16 au) in 2018 to (nominally) just 6.9-million kilometers (0.046 au) in 2025. During this period the field of view of the WISPR camera will evolve accordingly, recording white-light observations of the solar corona and solar outflows with its two overlapping detectors, WISPR-Inner and WISPR-Outer (hereafter, WISPR-I and WISPR-O). Both telescopes record white light observations with a 1920$\times$2048 pixel Advanced Pixel Sensor (APS) detector, though almost all images are downlinked at half resolution. The WISPR-I telescope has a bandpass of 490--740 nm, a plate scale of 1.2 arcmin/pixel, and an effective field of view of 13.5$^{\circ}$ - 53$^{\circ}$ ($^{\circ}$elongation). The WISPR-O telescope has a bandpass of 475--725 nm, a plate scale of 1.7 arcmin/pixel, and an effective field of view of 53$^{\circ}$ - 108$^{\circ}$ ($^{\circ}$elongation).

In Table~\ref{tab:enc-overview} we provide a summary of the observing circumstances for the WISPR-I and WISPR-O cameras during the spacecraft's first nine solar encounters, including the effective bounds of the fields of view and the minimum distance between the spacecraft and the orbit of Phaethon. The first column lists the encounter number. The second column provides the dates for the nominal encounter mission (though limited data are obtained outside of this range). The third column gives the perihelion distance of \textit{PSP} in astronomical units (au). The fourth and fifth columns provide the effective field of view of WISPR-I and WISPR-O at perihelion, in units of solar radii (R$_\odot$, where 1~R$_{\odot} = 695,000~\mathrm{km} = 0.00465$~au). The sixth column provides the minimum approach distance between \textit{PSP} and the orbit of Phaethon, and the date and time of that close approach. The seventh column provides the true anomaly of Phaethon itself ($\nu$\textsubscript{Phaethon}) at the time of \textit{PSP}'s perihelion. It should be noted that \textit{PSP}'s proximity \textit{to the observed dust trail} cannot be determined as we do not know the structure/extent of the trail itself; the provided \textit{PSP}-Phaethon orbit distance is thus an approximation only for this unknown distance, and any future reference to the ``PSP-trail'' distance shares this assumption.

\begin{table}[ht!]
\centering
\begin{tabular}{ |p{0.5cm}||p{2cm}|p{2cm}|p{1.7cm}|p{1.7cm}||p{3.8cm}|p{2.4cm}|  }
    \hline
    \multicolumn{7}{|c|}{PSP/WISPR Encounters 1 - 9 overview} \\
    \hline
    Enc. & Obs. Dates & \textit{PSP} Perihelion (au) & WISPR-I FOV (R$_{\odot}$) & WISPR-O FOV (R$_{\odot}$) & 3200-PSP Min. Dist. (au) \newline and Date (UT) & $\nu$\textsubscript{Phaethon} at \newline \textit{PSP} perihelion \\
    \hline
    1 & 2018 Oct 31 - \newline 2018 Nov 11 & 0.166 & 8 -- 32 & 30 -- 65 & 0.0765 \newline Nov 07 17:44 & 182$^{\circ}$\\
    2 & 2019 Mar 30 - \newline 2019 Apr 10 & 0.166 & 8 -- 32 & 30 -- 65 & 0.0765 \newline Apr 06 12:56 & 200$^{\circ}$\\
    3 & 2019 Aug 27 - \newline 2019 Sep 07 & 0.166 & 8 -- 32 & 30 -- 65 & 0.0765 \newline Sep 03 08:07 & 154$^{\circ}$\\
    4 & 2020 Jan 23 - \newline 2020 Feb 04 & 0.130 & 6 -- 25 & 23 -- 51 & 0.0563 \newline Jan 31 19:51 & 176$^{\circ}$\\
    5 & 2020 Jun 01 - \newline 2020 Jun 13 & 0.130 & 6 -- 25 & 23 -- 51 & 0.0563 \newline Jun 09 18:37 & 187$^{\circ}$\\
    6 & 2020 Sep 21 - \newline 2020 Oct 02 & 0.095 & 5 -- 18 & 17 -- 37 & 0.0387 \newline Sep 30 00:18 & 204$^{\circ}$\\
    7 & 2021 Jan 12 - \newline 2021 Jan 23 & 0.095 & 5 -- 18 & 17 -- 37 & 0.0387 \newline Jan 20 08:42 & 148$^{\circ}$\\
    8 & 2021 Apr 24 - \newline 2021 May 04 & 0.074 & 4 -- 14 & 13 -- 29 & 0.0277 \newline May 01 22:31 & 169$^{\circ}$\\
    9 & 2021 Aug 04 - \newline 2021 Aug 15 & 0.074 & 4 -- 14 & 13 -- 29 & 0.0277 \newline Aug 12 08:56 & 179$^{\circ}$ \\
    \hline
\end{tabular}
\caption{Specifications of the first nine \textit{PSP}/WISPR solar encounter periods. The second column provides the dates for the nominal encounter mission (though limited data are often obtained outside of this range). The third column gives the perihelion distance of \textit{PSP} in astronomical units (au) during that orbit. The fourth and fifth columns provide the effective field of view of WISPR-I and WISPR-O at perihelion, in units of solar radii (R$_\odot$, where 1~R$_{\odot} = 695,000~\mathrm{km} = 0.00465$~au). The sixth column provides the minimum approach distance between \textit{PSP} and the orbit of 3200 Phaethon, and the date and time of that close approach. The seventh column provide the true anomaly of Phaethon itself ($\nu$\textsubscript{Phaethon}) at the time of \textit{PSP}'s perihelion.  \label{tab:enc-overview}}
\end{table}

\subsection{WISPR Observations}\label{sec:wispr-obs}

In each Encounter, the parameters for WISPR observations (e.g., exposure times, image size, cadence) have varied in response to a number of factors including available telemetry and anticipated coronal brightness, as well as being modified to enhance the anticipated science return. In particular, as \textit{PSP} has approached closer to the Sun, the brightness in the inner portion of the field has increased substantially, requiring experimentation and subsequent reduction of exposure times to avoid saturation. These issues are discussed in more detail by \cite{Hess2021}. Here, as opposed to explicitly stating the observing parameters for all nine encounters thus far, we will instead refer to that publication and the WISPR project website\footnote{https://wispr.nrl.navy.mil/wisprdata}, from which so-called `summary files' -- plain text metadata for all recorded observations -- are readily available.

In Paper I we studied observations recorded by WISPR-I on 2018 November 5, during \textit{PSP}'s first solar encounter, noting that a small portion of the dust trail was visible in WISPR-O, but was not considered at that time due to ongoing calibrations regarding the WISPR-O data. Since that time, the calibration procedures for WISPR-O have been finalised and published \citep{Hess2021} and a Level-3 (L3) data product for WISPR-O is routinely produced, thus enabling us to extend our investigation to this camera. This is fortuitous as the dust trail is now better observed in WISPR-O than WISPR-I, primarily because the evolving orbit of \textit{PSP} has shifted the timing of our close approach to the trail to coincide with the trail's passage through WISPR-O, particularly since Encounter 4. Accordingly, the emphasis of our analyses of the trail has shifted primarily towards the WISPR-O but limited observations from WISPR-I are considered.

In this study, we examine fully-calibrated L2 and L3 observations from primarily WISPR-O for the first nine \textit{PSP} encounters, hereafter referred to as E1 through E9. However, we note here that observations in E2 and E3 were of lower quality due to early-mission operational and instrumental problems, and thus we do not consider data from those encounters here. All WISPR data products (L1, L2 and L3) are publicly available on the WISPR instrument website and detailed extensively by \cite{Hess2021}. The process for producing and calibrating L3 data is outlined in Section 2 of \cite{Battams2020}. Thus we will omit detail here, but briefly: the L3 data product is the result of subtracting a calibrated background image (L2b) from the Level-2 (L2) data product. The background subtraction removes stationary features in the scene of the WISPR image, namely the dominant F-corona background along with some of the faint and diffuse K-corona component, stray light patterns of instrumental origin, and excess brightness due to circumsolar dust features that appears ``fixed'' in the field of view (FOV) due to the orbital characteristics of the S/C \citep[e.g., the dust ring nearby to Venus'orbit;][]{Stenborg2021VenusRing}. A number of different image processing algorithms were applied to the L2 and L3 data, depending on the nature of the analysis in question; these algorithms are detailed with examples in Appendix~\ref{app:A1}.

\subsection{Dust Trail Observations by Encounter}

The nature of \textit{PSP}'s orbit is such that certain Encounters share almost identical orbits and can be grouped together accordingly. That is, during E1, E2, and E3, the spacecraft followed a largely identical orbit, and consequently our field of view was very similar in each of these encounters. The same situation applied to E4 -- E5, E6 -- E7, and E8 -- E9, respectively, with Venus fly-by gravity assists placing the spacecraft into new orbits after E3, E5, and E7, and data obtained within each of those orbital configurations offering essentially the same field of view. Thus the WISPR view of the orbit of Phaethon \textit{within each of the orbital configurations} remained almost identical, but the transition to each new orbit configuration presented a slightly different view of the Phaethon orbit (and thus the dust trail), as described in Table~\ref{tab:enc-overview}.

\begin{figure}[ht!]
\centering
\includegraphics[width=0.9\textwidth]{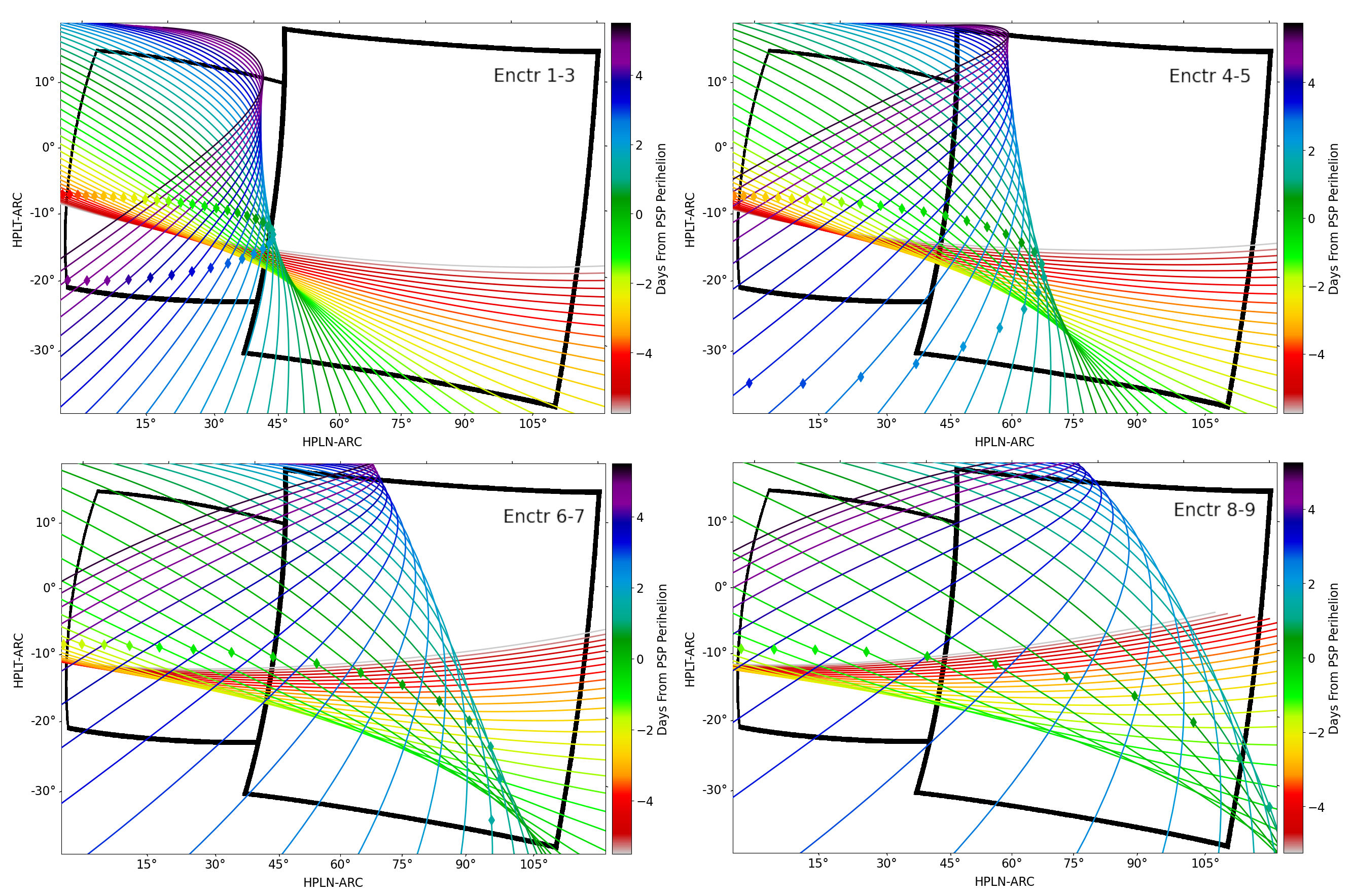}

\caption{Summary graphics showing the trajectory of Phaethon's orbit across the WISRP-I (smaller black outline, left) and WISPR-O (larger black outline, right) for each orbital configuration, with the fields of view projected into a helio-projective coordinate system. The color encoded into each line references the time in days from \textit{PSP}'s perihelion point in its orbit. The diamond shapes indicate the location of Phaethon's perihelion point along those lines.}
\label{fig:enctr-summary-time}
\end{figure}

\begin{figure}[ht!]
\centering
\includegraphics[width=0.9\textwidth]{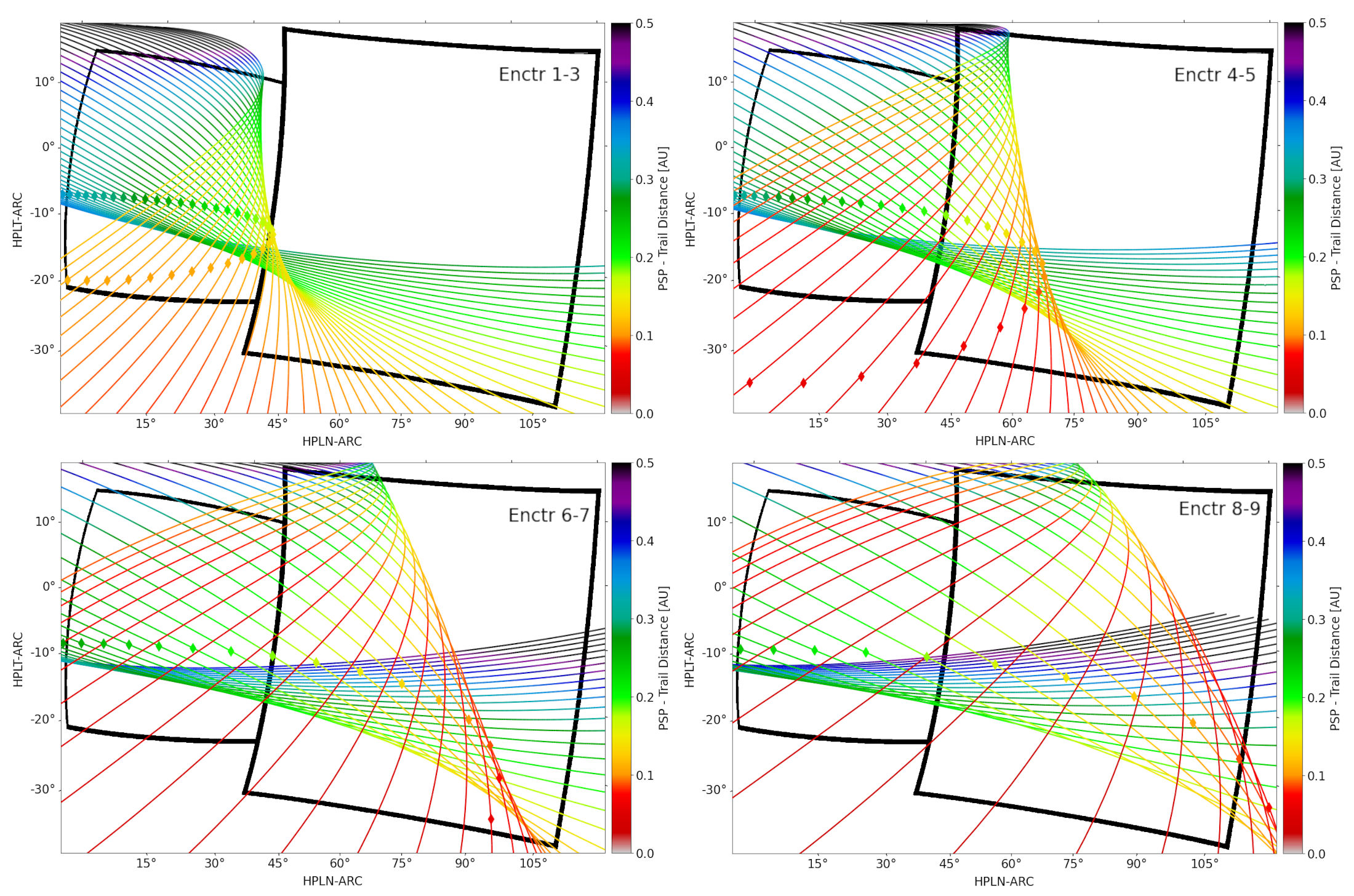}

\caption{As Figure~\ref{fig:enctr-summary-time}, but with the color encoding on each line referencing the physical distance (in au) between \textit{PSP} and that portion of the Phaethon orbit. }
\label{fig:enctr-summary-dist}
\end{figure}

\begin{figure}[ht!]
\centering
\includegraphics[width=0.9\textwidth]{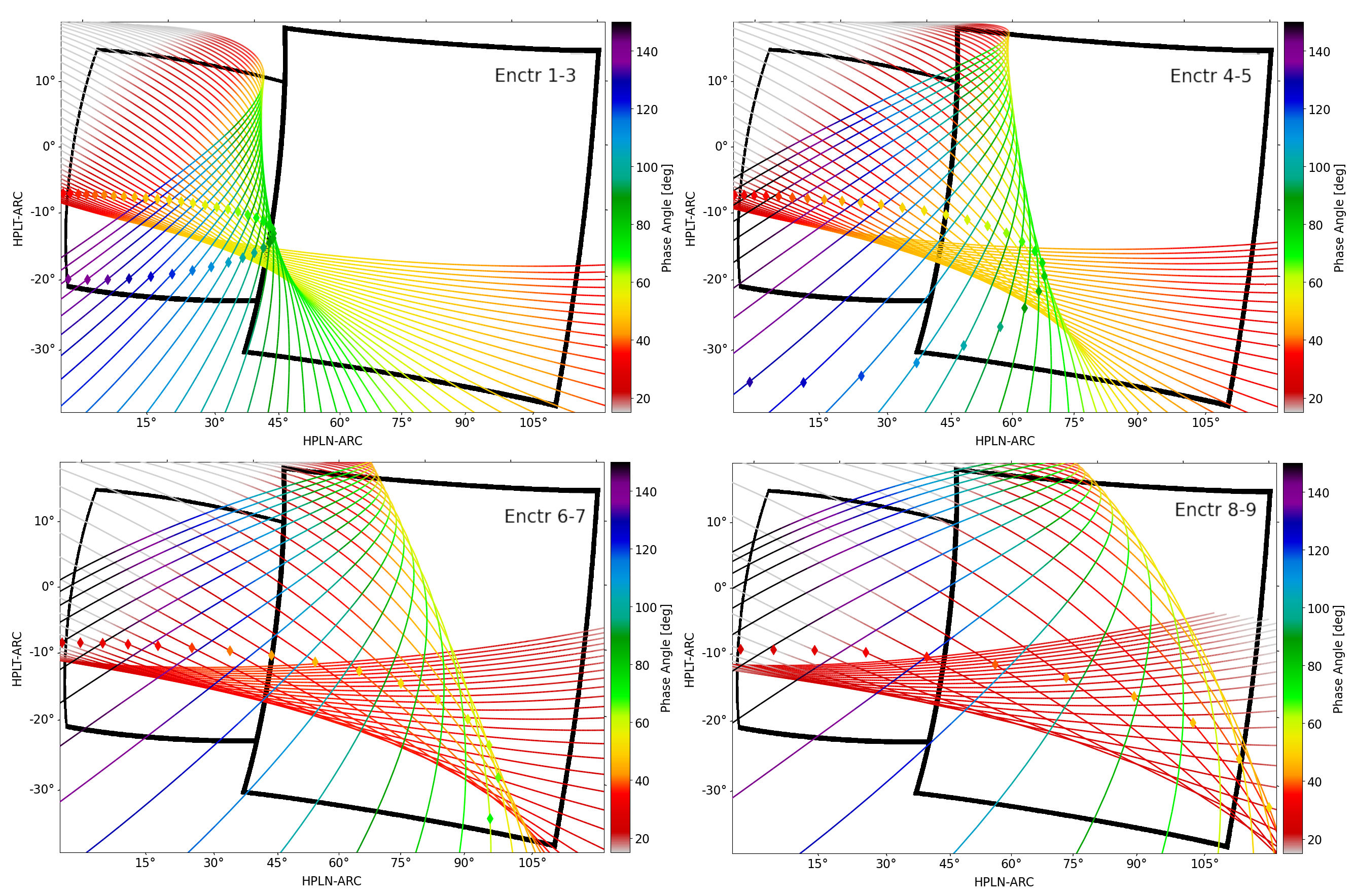}

\caption{As Figure~\ref{fig:enctr-summary-time}, but with the color encoding on each line referencing the phase angle of our observation of the Phaethon orbit at each point along the line, where the phase angle is the angle between the PSP-Sun and Phaethon-Sun lines.}
\label{fig:enctr-summary-phase}
\end{figure}

The following four-panel Figures, \ref{fig:enctr-summary-time} through \ref{fig:enctr-summary-phase}, provide a graphical summary of the orientation and evolution of Phaethon's orbit in both the WISPR-I and WISPR-O fields of view through the first nine encounters, with one panel per orbit configuration. These Figures use a solid black border to indicate the edges of the fields of view of WISPR-I and WISPR-O, projected into a helio-projective coordinate system as indicated by the axes (these values are only approximate as slight pointing changes in each encounter vary the longitude by up to $\sim$3$^\circ$, and the latitude by $<0.1^\circ$). In all cases, the Sun is outside the FOV, to the far left (see also Figure 4 of Paper I). The color lines in these Figures indicate the orientation of Phaethon's orbit projected through the fields of view with six-hour time steps between each line. The colors encoded into each line are described as follows, and the diamond symbols indicate the perihelion point of Phaethon's orbit.

Figure~\ref{fig:enctr-summary-time} shows the orbit of Phaethon in each of the four \textit{PSP} orbital regimes thus far with color used to indicate elapsed time from \textit{PSP}'s perihelion. Each individual line is a single color representing the orientation of the orbital path at that timestamp. It is by fortunate coincidence that \textit{PSP}'s perihelion always falls roughly mid-way through the period in which Phaethon's orbit is in the field of view, providing a useful reference point here. The primary purpose of this Figure is to demonstrate how the trail orientation changes over time, both within each encounter and across all four orbital configurations, and to illustrate how the trail has moved further into the larger WISPR-O FOV in later encounters. The relative speed with which the trail passes through the field of view can also be inferred here.

In Figure~\ref{fig:enctr-summary-dist} the color encoding references the distance between Phaethon's orbit and \textit{PSP} at each time step. From this Figure we can see that the point of closest approach between \textit{PSP} and the Phaethon orbit (dust trail) again occurs increasingly in the WISPR-O FOV, particularly for E6 through E9.
Figure~\ref{fig:enctr-summary-phase} again presents the same information but with the phase angle of the trail indicated by the colorbar. We will note here that the trail phase angles throughout all encounters remain broadly in the range $20^{\circ}-140^\circ$ which, assuming the dust to behave like typical cometary dust, would not contribute substantially to the enhancement/reduction in apparent trail brightness \citep[][]{Schleicher2011}. (This assumption, which may not be entirely valid, is discussed in more detail later in the manuscript.)

\begin{figure}[ht!]
\centering
\includegraphics[width=0.9\textwidth]{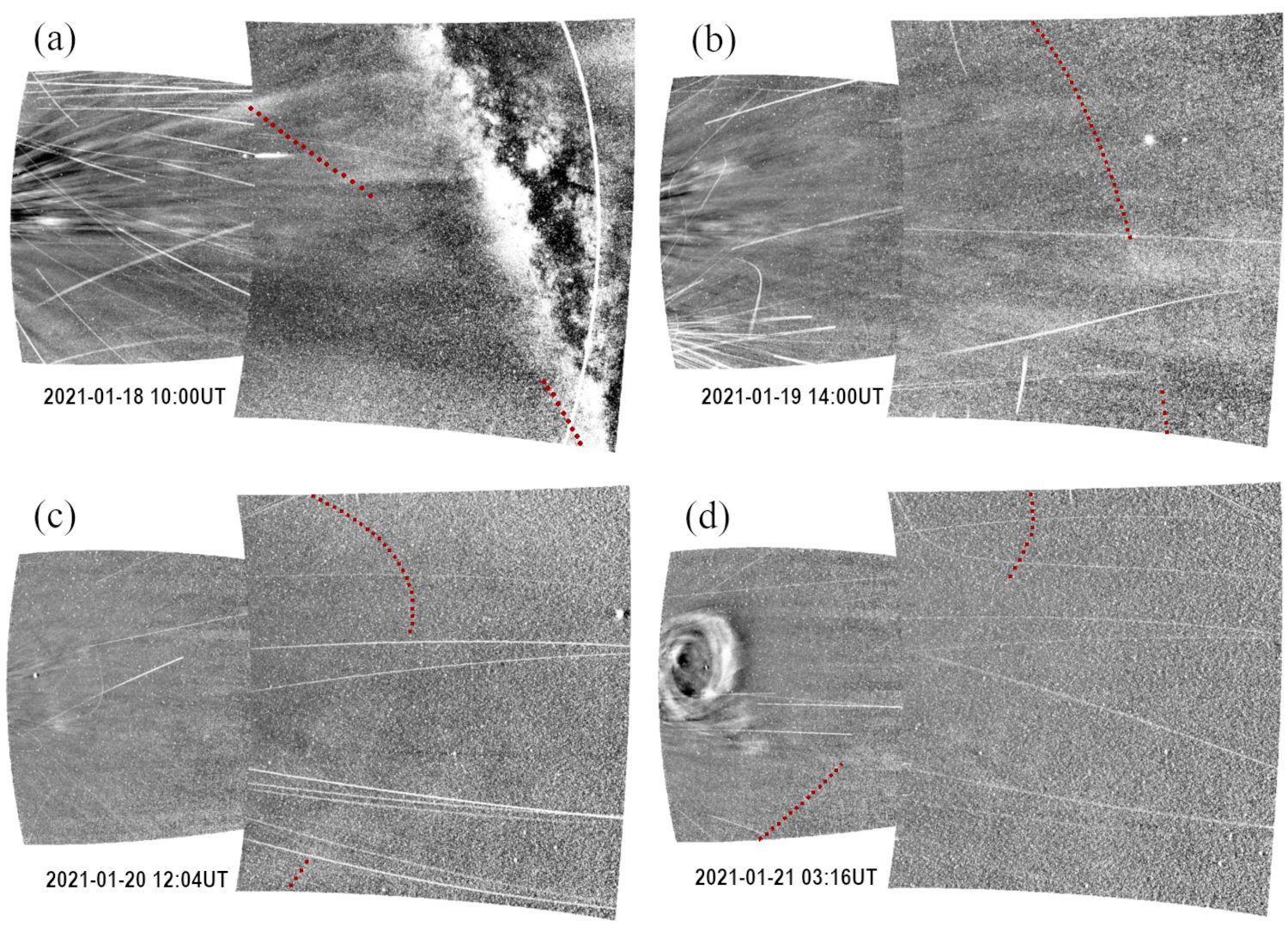}

\caption{A snapshot of four WISPR-I/WISPR-O composite observations recorded during E7, at the date and time indicated on each panel, and enhanced with the LW processing algorithm. On each panel, a dashed line has been placed over a portion of the orbit of Phaethon to indicate its path through the fields of view, with the central part of that dashed line omitted so that the trail can be seen more clearly. Numerous impact ejecta streaks/arcs can be seen in all four images, and a coronal mass ejection in the WISPR-I image on panel (d)}
\label{fig:e7-summary-images}
\end{figure}

These figures are a valuable guide for summarizing the viewing circumstances of the dust trail (Phaethon's orbit) at any point during the first nine encounters. For example, the earliest trail appearance in Encounters 1-3, at -5 days from \textit{PSP}'s perihelion, is nearly horizontal through both FOVs, with a nearly constant distance from \textit{PSP} of 0.2-0.3 au. The phase angle is at about 30$^\circ$ at the outer edge of WISPR-O, peaks at about 50$^\circ$ near the right (outer) edge of WISPR-I, and then decreases to $\sim$30$^\circ$ at the left (inner) edge of WISPR-I. Figures~\ref{fig:enctr-summary-time} through \ref{fig:enctr-summary-phase} also highlight how the successive orbit changes have moved the closest portions of the trail further into the WISPR-O FOV, and broadly speaking it is at these times in WISPR-O that the trail is most clearly visible in the data. However, the visibility of the trail across all four orbit configurations, and even from encounter to encounter within a given configuration, was highly variable for a number of reasons. To further illustrate the variable viewing geometry of the trail, Figure~\ref{fig:e7-summary-images} displays a composite of WISPR observations as processed with the so-called ``LW'' algorithm (described in Appendix~\ref{app:A1}) at four time instances during E7, where we can observe the dust trail (partly indicated with the broken red dotted line) with different orientations as the encounter progressed.

Primarily, the trail was consistently at, or at least barely above, noise-levels in the data. This meant that any variation in the background scene, whether from planets, stars, the Milky Way (Figure~\ref{fig:e7-summary-images}(a)), coronal streamers, outflows (Figure~\ref{fig:e7-summary-images}(b)), or coronal mass ejections (far left of Figure~\ref{fig:e7-summary-images}(d)), can easily overwhelm the dust trail signal. Furthermore, \textit{PSP} experiences a very high rate of micro-meteoroid impacts, many of which result in a spray of bright ejecta particles crossing the WISPR fields of view \citep[][]{Malaspina2022}. In extreme circumstances, these events can completely blind the instrument, and even at low levels we observe at least a few impact ejecta streaks in almost every WISPR image, as seen in all four panels of Figure~\ref{fig:e7-summary-images}.

In the following Sections we provide analyses of these data focusing on two separate (though complementary) aspects of the observations: morphology and physical properties. For the former, we will look specifically at analyses of the cross-section profile of the dust trail in all four orbit configurations, and highlight a key new result regarding the positioning of the dust trail relative to Phaethon's orbit. For the latter, we will look at the photometric properties of the dust trail, providing an update to the results presented in Paper I, and investigating further into the estimated mass of the dust observed by WISPR.

\section{Analysis} \label{sec:analysis}

\subsection{Dust Trail Morphology and Structure} \label{sec:morph}

\begin{figure}[ht!]
\centering
\includegraphics[width=0.75\textwidth]{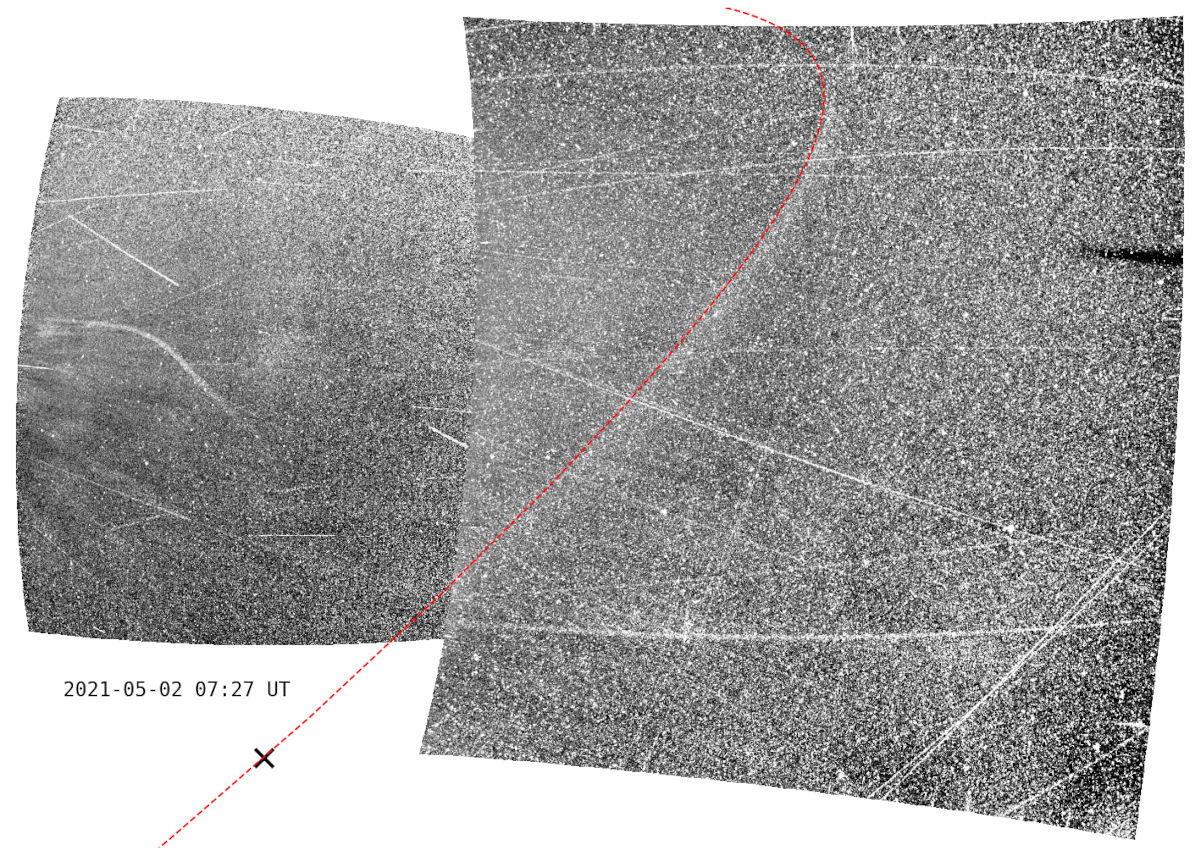}

\caption{WISPR-I and WISPR-O composite recorded on 2021 May 02 (E8). Phaethon's orbit is indicated by the dashed red line, with a black X indicating the closest point of Phaethon's orbit to \textit{PSP} at the time of observation, which happened to be outside of the field of view. The dust trail can be seen relatively clearly following Phaethon's orbit, though appearing offset largely anti-sunward. These data were processed using the LW algorithm.}
\label{fig:trail-offset}
\end{figure}

The increasing proximity of \textit{PSP} to the Phaethon orbit and the associated dust trail, particularly in E6--E9, has provided ever-improving views of the trail both in terms of the apparent width of the trail, as well as the clarity with which it can be observed relative to instrumental noise. Most strikingly, these latest encounters revealed that the dust trail is apparently \textit{not} centered on the orbit of Phaethon as initially indicated by E1--E2 results presented in Paper~I. This is highlighted by Figure~\ref{fig:trail-offset}, which presents a composite of WISPR-I and WISPR-O recorded on 2021 May 02 (E8), processed with the LW algorithm. The red dashed line indicates the orbit of Phaethon (using the 2021 orbit solution for Phaethon) and a black cross (outside of the instrument field of view) indicates the point of minimum distance between \textit{PSP} and Phaethon's orbit. In these observations, and many others of its kind, the dust trail appears to lie on the anti-sunward side of the Phaethon orbit, with minimal apparent overlap between the orbit and the dust trail. At no point was an obvious \textit{separation} observed between the dust and Phaethon's orbit, and at certain times in each encounter there was no apparent offset at all. In E1--E5, the offset was not visually detectable at any time, though as demonstrated later was indeed present.

Given the multiple factors at play in \textit{PSP}'s dynamic and rapidly evolving fields of view, we focused on examining trends in the \textit{trail offset as a function of the true anomaly} of the portion of Phaethon's trail in question. That is, within each orbit configuration we would identify sequences of images in which the trail was at least reasonably visible with few major dust/noise artefacts inhibiting the view. We would then select a central portion of the trail (Phaethon orbit) based on its helio-projective cartesian (HPC) coordinates and produce a Gaussian fit to a cross-trail profile, as described in Appendix~\ref{app:A}. This provided an estimate of the angular offset of the Gaussian peak from Phaethon's orbit, which we could convert to a physical separation based on the distance between \textit{PSP} and the targeted location along Phaethon's orbit. Multiple points in each observation were selected, and the same true anomaly location was also selected in multiple observations, to get as broad a survey as the data would allow.

The faintness of the trail signal meant that simple cross-trail intensity profiles extracted from even the clearest observations in unprocessed L3 data resulted in a noisy line, with no central peak evident. Thus we developed a stacking technique, appropriate for the somewhat complex WISPR image scene, that allowed us to extract relatively clear cross-trail profiles and model them with simple Gaussians to record the pertinent properties. This technique, explained in detail with examples in ~\ref{app:A}, allowed us to routinely estimate trail offset values, trail widths and, in certain circumstances (noted in the Appendix) provide alternative means for photometric estimates. However, we note briefly here that while the trail offsets that we report are clearly real, despite the large error bars, we cannot be so confident in the width of the Gaussians, which were very much influenced by the passage of background solar outflows, and even less confident in the Gaussian amplitudes, which vary significantly as a function of the baseline intensity in any given portion of the field of view (i.e., pixels near the Sun are substantially brighter than those far from the Sun).

\begin{figure}[ht!]
\centering
\includegraphics[width=0.7\textwidth]{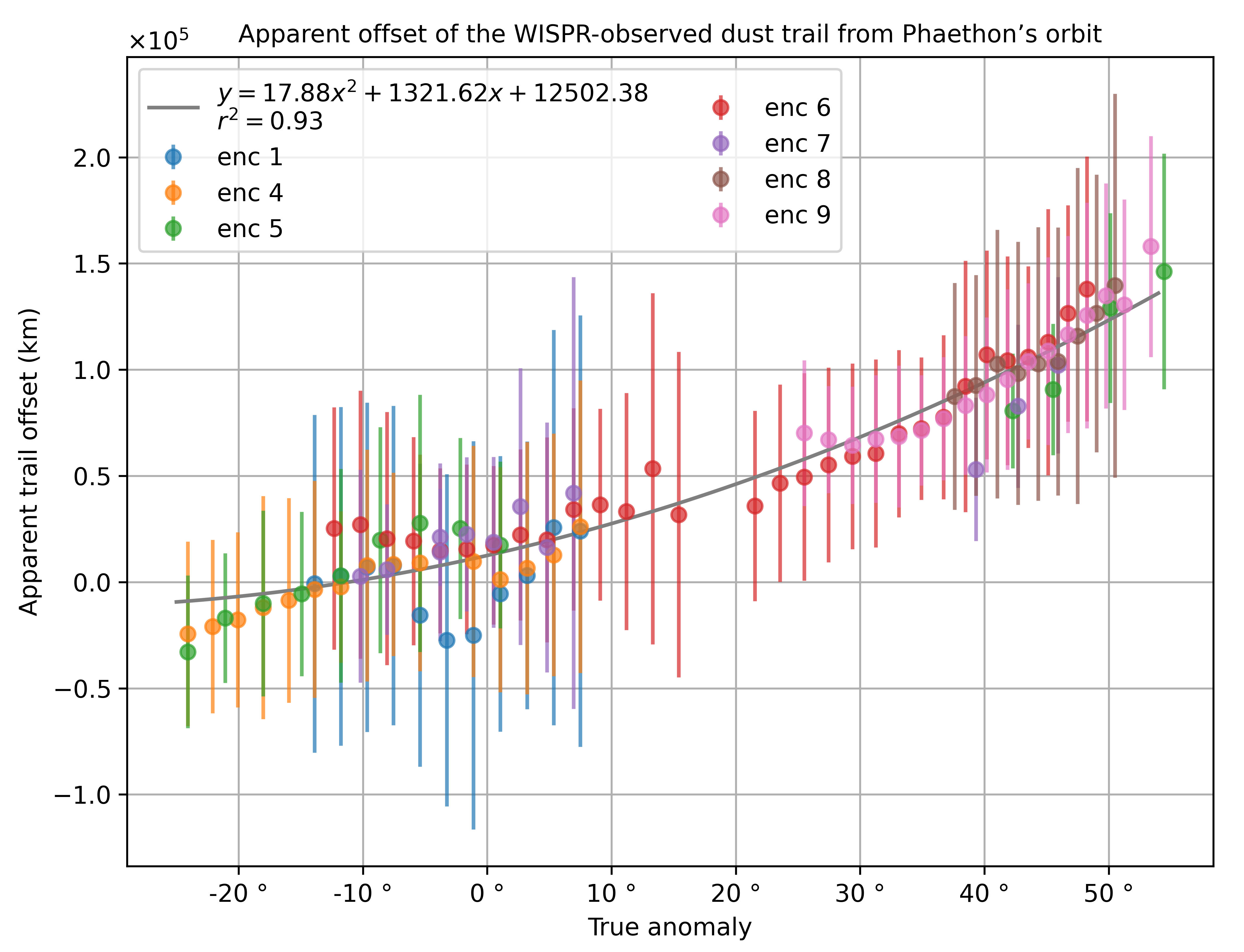}

\caption{Plot of the apparent physical offset (in $10^5$ kilometers) of the dust trail from Phaethon's orbit as a function of true anomaly (degrees), based on the central peak locations of Gaussians fitted to cross-trail profiles obtained from various points throughout most of the first nine encounters. A second order polynomial fit is shown with the fit parameters in the Figure legend along with the $r^2$ value (coefficient of determination, which is the square of the correlation coefficient, \textit{R}) for the polynomial, but we note that this polynomial and its coefficients are not intended to be physically meaningful.}
\label{fig:offset-plots}
\end{figure}

Figure~\ref{fig:offset-plots} displays the results of a survey of E1 and E4--E9 observations, plotting angular offset converted to a true apparent physical distance ($10^5$~km) as a function of the True Anomaly (degrees) of the observation point. The error bars are formed from the standard deviations of multiple profiles fitted to the same portion of the orbit in different sized regions of interest. The points have a high degree of scatter due to the excessive noise in the observations causing poor Gaussian fits, and the E1 points show anomalous values that could be due to either the lower quality observations at that time, or a slightly different viewing geometry of the trail. Nonetheless, the results unambiguously show a divergence of the dust trail and Phaethon's orbit as a function of true anomaly. This leads us to a key result of this study: namely, that \textit{PSP} is \textit{not} observing a dust trail that perfectly follows Phaethon's orbit, but instead is viewing presumably Phaethon-related dust that has either evolved out of that orbit or was never released into an orbit identical to that of Phaethon. This result is discussed further in Section~\ref{sec:discuss}.

\subsection{Dust Trail Photometry} \label{sec:photom}

In Paper I we determined the visual (V) magnitude of the trail to be 15.8~$\pm$~0.3 per pixel with a surface brightness of 25.0 mag arcsec$^{-2}$. In that study, we noted the challenge of performing aperture photometry on the narrow, low-signal trail in such high-noise observations. This situation has improved slightly in recent encounters, with the technique employed in the initial investigation working somewhat more reliably. This technique is essentially a naive background-subtraction method in which small (e.g. 4$\times$4 pixel) regions of the trail are considered, with a suitable ``background'' images being derived from averaging four temporally adjacent images of the same sky region. This was still challenging as the trail moved across the sky at different rates depending on \textit{PSP}'s location and velocity at the time, and the constant noise sources added extra complication.

Nonetheless, this technique was applied to several images in each encounter, and along different parts of the trail representing different True Anomalies (and heliocentric distances). Unlike the first Encounter, in which the trail was best seen in WISPR-I, all photometric analyses here were based on WISPR-O observations. Using the latest photometric calibration coefficients (\textit{P. Hess, Private Comm.}), we consistently found per-pixel visual magnitudes varying from $15.9-16.3$, with no apparent trends as a function of True Anomaly and/or heliocentric distance. Based on these analyses we find an average per-pixel visual magnitude value for the trail of 16.1$\pm$0.3 per pixel. This is within the error bars of the estimates published in our initial study (15.8$\pm$0.3 per pixel), albeit slightly fainter on average than the values obtained in that study.

In the same vein, a useful facet of the trail profile stacking technique was that it could be used to provide a convenient alternative method to estimate the photometry of the trail simply by observing the height of the Gaussian peak above the (highly variable) noise floor of the cross-trail profile, as detailed in Appendix~\ref{app:A-photom}. We did not apply this methodology rigorously to a broad swath of the observations, instead relying primarily on visually inspecting the data to ensure consistency. As expected, the data were very noisy, but the background-peak signal routinely fell within $\sim$0.3$\times$10$^{-14}$ $B_{\sun}$ and $\sim$0.7$\times$10$^{-14}$ $B_\sun$, corresponding to visual (V) magnitudes of 16.7 and 15.7 respectively. Again, this was entirely in-line with our slightly more consistent aperture photometry results of V$\sim$16.1$\pm$0.3 per pixel. With the specifics of this value not particularly impactful on our bigger picture conclusions, we hereafter accept V$\sim$16.1$\pm$0.3 per pixel as the visual magnitude of the trail without further analysis. This corresponds to a surface brightness of 26.1~mag~arcsec$^{-2}$, which remains comparable to the 25.0~mag~arcsec$^{-2}$ value determined in Paper I, albeit somewhat fainter, and still upholds all related conclusions presented in that study.

\subsection{Dust Mass Estimates} \label{sec:mass}

In Section 3.3 of Paper I we detailed our methodology and assumptions in estimating a total mass of dust in the trail to be $\sim(0.4-1.3){\times}10^{12}$~kg, which at the time we noted as being somewhat less than, but largely congruent with, estimates for the total mass of the Geminid stream \citep[e.g.][]{Ryabova2017}. These results were based almost entirely on WISPR-I E1 observations, with the spacecraft relatively distant from the trail ($\sim$0.17 au at closest approach), and the data quality comparatively poor. We can now reassess these mass estimates using a broader range of parameters derived from all nine encounters, and at different viewpoints relative to the orbit of the trail. In all calculations, our methodology and assumptions are identical to that in Paper I, based on \citet{Jewitt1991}'s reformulation of \citet{Russell1916}'s equation to estimate the cross section of dust as
\begin{equation}
    C_d = \frac{2.25{\times}10^{22}{\pi}r_\mathrm{H}^2{\Delta}^2}{p_V{\phi}({\alpha})}10^{0.4(m_{\odot} - m_{phaethon})}
\end{equation}
where $C_{d}$ is the dust cross section in m$^2$, $r_\mathrm{H}$ is the heliocentric distance of the dust trail in au, $\Delta$ is the {\it PSP}-trail distance in au, $p_V$ the geometric albedo,  $\alpha$ is the phase angle, $m_{\odot}$ is the apparent $V$ magnitude of the Sun ($-$26.74) and $m_{phaethon}$ is the estimated $V$ magnitude of the dust trail. Then we can use
\begin{equation}\label{eq:mass}
    m_{d} = \frac{4}{3}{{\rho}_d}{C_d}\bar{a}
\end{equation}
with ${\rho}_d = 2900$~kg~m$^{-3}$ (the bulk density of grains in the trail; \citealt{Babadzhanov2009}) and $\bar{a}$ as the dust grain radius (mm) to estimate the total mass of dust in the trail.

In Paper I we assumed the radius of the dust grains to be 0.5~mm -- a value based upon the typical grain sizes observed within the Geminid stream. We have no knowledge of the grain sizes of the trail observed by WISPR thus, absent a significant modeling exercise, we will again nominally assume that same grain size but note that the total mass scales linearly with dust grain size, per Equation~\ref{eq:mass}.

We determined values for $r_\mathrm{H}$, $\Delta$, $\alpha$, and the estimated trail width, at 495 points across all encounters except E2 and E3, using WISPR-I observations for E1 and WISPR-O for all other Encounters. For each observation $\alpha$, we made two different evaluations of the phase function (${\phi}(\alpha)$): (i) via the {\it HG} parameters for Phaethon given in \citet{tabeshian19}, for which we assume the dust grains behave like `mini-Phaethons'; and (ii) using the phase function for cometary dust as determined by Schleicher and Marcus\footnote{\url{https://asteroid.lowell.edu/comet/dustphase_details.html}}, which assumes the dust to be more like cometary dust. The results of these calculations are shown in Table~\ref{tab:mass-calcs} with values for each individual encounter and an average across all encounters. We note here that while precision is stated to one decimal place, the broader uncertainties on these estimates will outweigh this implied level of precision, and these values should be regarded as approximations only.

\begin{table}[ht!]
\centering
\begin{tabular}{ |p{1.0cm}||p{2cm}|p{3cm} | p{3cm} |}
    \hline
    \multicolumn{4}{|c|}{Dust Trail Mass Estimates (kg)} \\
    \hline
    Enc. & Num. Points & `Mini-Phaethons' & Cometary Grains  \\
    \hline
    1 & 75 & (2.6 -- 3.0) ${\times}10^{11}$ & (7.0 -- 7.3) ${\times}10^{10}$\\
    4 & 39 & (1.8 -- 2.1) ${\times}10^{11}$ & (3.6 -- 4.6) ${\times}10^{10}$\\
    5 & 51 & (1.3 -- 4.5) ${\times}10^{11}$ & (3.3 -- 4.7) ${\times}10^{10}$\\
    6 & 130 & (0.6 -- 3.9) ${\times}10^{11}$& (2.3 -- 5.3) ${\times}10^{10}$\\
    7 & 56 & (1.4 -- 4.6) ${\times}10^{11}$ & (2.3 -- 3.1) ${\times}10^{10}$\\
    8 & 63 & (2.1 -- 5.0) ${\times}10^{11}$ & (1.9 -- 3.5) ${\times}10^{10}$\\
    9 & 81 & (2.0 -- 3.8) ${\times}10^{11}$ & (1.4 -- 2.4) ${\times}10^{10}$\\
    \hline
    All & 495 & (1.4 -- 4.1) ${\times}10^{11}$ & (1.7 -- 5.3) ${\times}10^{10}$\\
    \hline
\end{tabular}
\caption{Estimates of the mass of the dust trail based on multiple observations throughout seven of the nine encounters studied. The mass ranges are based on one standard deviation of the mean masses, and the final (``All'') value is derived from all results collectively. Column three assumes the dust grains' phase function to behave like that of Phaethon (i.e. as `mini-Phaethons'), and the fourth column for the phase function to behave like that of cometary grains.\label{tab:mass-calcs}}
\end{table}

Averaging across all encounters and assuming 0.5~mm dust grains, we find an estimated mass of (1 -- 4)${\times}10^{11}$~kg under the mini-Phaethon assumption, and (2 -- 5)${\times}10^{10}$ under the cometary dust grain assumption. The equivalent values for 0.05~mm and 5~mm dust grains are therefore factors of ten higher and lower in both cases, respectively. Values determined from each individual encounter likewise produce similar estimates, with all uncertainty ranges overlapping. All of these mass estimates are largely concurrent, though the cometary dust grain assumption leads to an approximate factor of ten decrease in mass. These estimates are somewhat below the lowest end of our $\sim$(4~--~13)${\times}10^{11}$~kg mass estimate reported in Paper I, but our understanding of the data, and methodology for extracting information, has improved markedly since the initial publication and thus we lend more weight towards these new results. Nonetheless, given the uncertainties in our assumptions, these estimates provide a relatively consistent picture for the mass of the trail on the order of $10^{10}$~kg -- $10^{11}$~kg. This is discussed in depth in Section~\ref{sec:discuss_dust_mass}.

\section{Discussion} \label{sec:discuss}

\subsection{Trail Morphology and Origins}\label{disc:morph_orb_prop}

Perhaps the most significant result from this study is that demonstrated by Figure~\ref{fig:offset-plots}, which makes a compelling argument that the WISPR-observed dust trail \textit{does not share the same orbit as Phaethon}, as nominally suggested to be the case in Paper I. This finding was initially overlooked simply because the data used in that study contained insufficient information to discern any appreciable separation from Phaethon's orbit, and our techniques for handling the data were still under development.

We have investigated, and subsequently ruled out, the possibility that the apparent trail offset has any relationship to the angle at which WISPR is viewing the orbital plane of Phaethon. Using the JPL Horizons Ephemeris web application\footnote{https://ssd.jpl.nasa.gov/horizons/app.html}, we can see that during all nine encounters, \textit{PSP} is essentially embedded in the plane of Phaethon's orbit and rarely leaves it during these encounter periods. For example, in 2018 Nov 05 - Nov 08 (E1) the plane angle changed only from -0.16$^{\circ}$ to 1.32$^{\circ}$, with very little noticeable change in trail offset. In E6, the plane angle varied from just 1.76$^{\circ}$ to 1.11$^{\circ}$, the latter value corresponding to some of the largest observed trail offsets. When we examine all encounters, we find the plane angle barely strays outside the range $\pm$2.0$^{\circ}$, with most trail observations occurring within $\pm$1.0$^{\circ}$. However, despite the plane-observing angle not offering any enlightenment, it is noteworthy that all of the WISPR observations of the dust trail occur at these extremely low plane angles when the optical depth would be the largest -- indeed, this may be a crucial clue in understanding how or why WISPR is the only instrument to make these white-light detections.

We also note in Figure~\ref{fig:offset-plots} that the minimum point of the offset does not appear to occur at Phaethon's perihelion point (i.e., True Anomaly of 0$^{\circ}$) but instead appears to reach an apparent offset of $\sim$0~km around -10$^{\circ}$, implying the dust follows an orbit that only differs marginally from that of Phaethon. A number of different mechanisms could produce such differences depending on whether the trail has simply evolved over some long period of time through the various forces at play, or if the dust was ejected into a different orbit at its creation. A recent study by \cite{Ryabova2022} demonstrated the existence of a number of orbital resonances of the Geminids with Earth, Venus, and Jupiter, each of which is likely to produce discrete dust trails in the Geminid stream distant (and unobservable) from Earth. However, the study notes that the quantity of trapped particles is likely to be very small, and the predicted resonances do not appear to align with our observed WISPR dust trail.

\begin{figure}[ht!]
\centering
\includegraphics[width=0.8\textwidth]{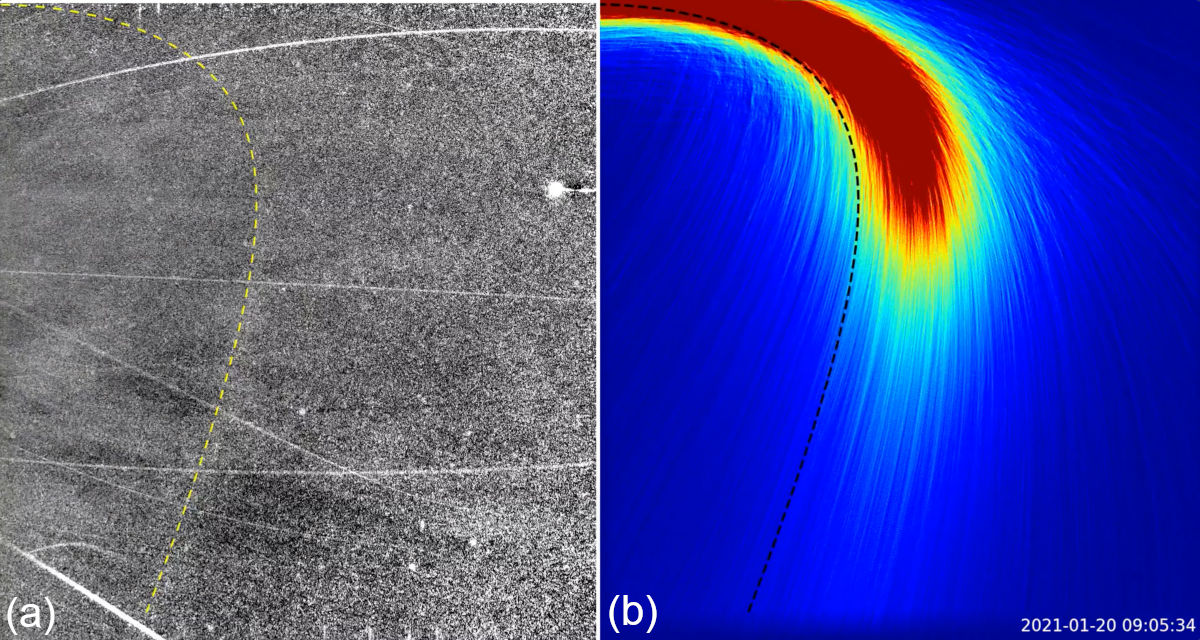}
\caption{Panel a: an LW processed WISPR-O image recorded on 2021 January 20 09:05UT, with a yellow dashed line indicating the orbit of Phaethon, with the dust trail very faintly following it. Panel b: a simulation showing $\sim$14,000 Geminid orbits integrated into the same field of view as panel a, with a black dashed line showing the orbit of Phaethon. The color scale of (b) represents the number of individual Geminid orbits that cross over any given pixel.}
\label{fig:geminids-e7}
\end{figure}

Regarding temporal evolution, we could nominally assume that the trail is some remnant from the event that created the Geminids. But this raises the question of how or why a discrete dust trail has remained close to the orbit of Phaethon whereas the bulk of the Phaethon-Geminid system appears widely distributed in space. In paper I we noted that visually the orbit of the dust did not appear to coincide well with that of the Earth-encountering Geminids. Figure~\ref{fig:geminids-e7} present a better illustration of this with panel (a) showing a processed WISPR-O image recorded on 2021 January 20 09:05UT with the dust trail seen faintly following the orbit of Phaethon (yellow dashed lines). Then Figure~\ref{fig:geminids-e7}(b) shows a simulation of approximately 14,000 Geminid orbits \citep{Jenniskens2018, Szalay2021} integrated onto a simulated version of panel (a), with a black dashed line showing the orbit of Phaethon. The color scale on (b) reflects the number of individual Geminid tracks that cross any given pixel in that field of view. We can see from this Figure (and have verified from other similar calculations) that the Earth-crossing portion of the shower does not overlap well with the trail observed by WISPR, and there is no visible evidence in the WISPR observations of the Earth-crossing Geminid stream. However, it is important to note that while the Geminids are seemingly concentrated around some central core or distribution peak in this simulation (i.e., the deep red portion of Figure~\ref{fig:geminids-e7}b), this peak is simply an indication of the sample bias with which we determine individual Geminid orbits. That is, these observations are based only upon meteors that cross Earth's orbit, whereas the ``true'' distribution of the stream could indeed be centered more closely to our observed dust feature. Results from \cite{Jones1986}, for example, found that Earth was likely some distance from the ``core'' of the stream, and may be gradually approaching it in the coming decades. Nonetheless, this still does not reconcile the trajectory and properties of the WISPR-observed dust trail with those of the Geminids.

Related to this, we can also raise the question of how the Earth-encountering Geminids have a demonstrated broad mass distribution from millimeter dust to reported 5~kg masses \citep{Yanagisawa2008}, yet WISPR is observing a discrete trail of some presumably consistent dust mass/size, with that dust having remained close to the parent despite most of the mass in the system dispersing far from Phaethon. Of course, studies such as \cite{Jones1986} dating back more than thirty years have commented on the relatively slow evolution of the trail, noting that the dispersion of the Geminids is primarily a result of gravitational perturbations rather than radiation pressure or Poynting-Robertson effects.

\begin{figure}[ht!]
\centering
\includegraphics[width=0.95\textwidth]{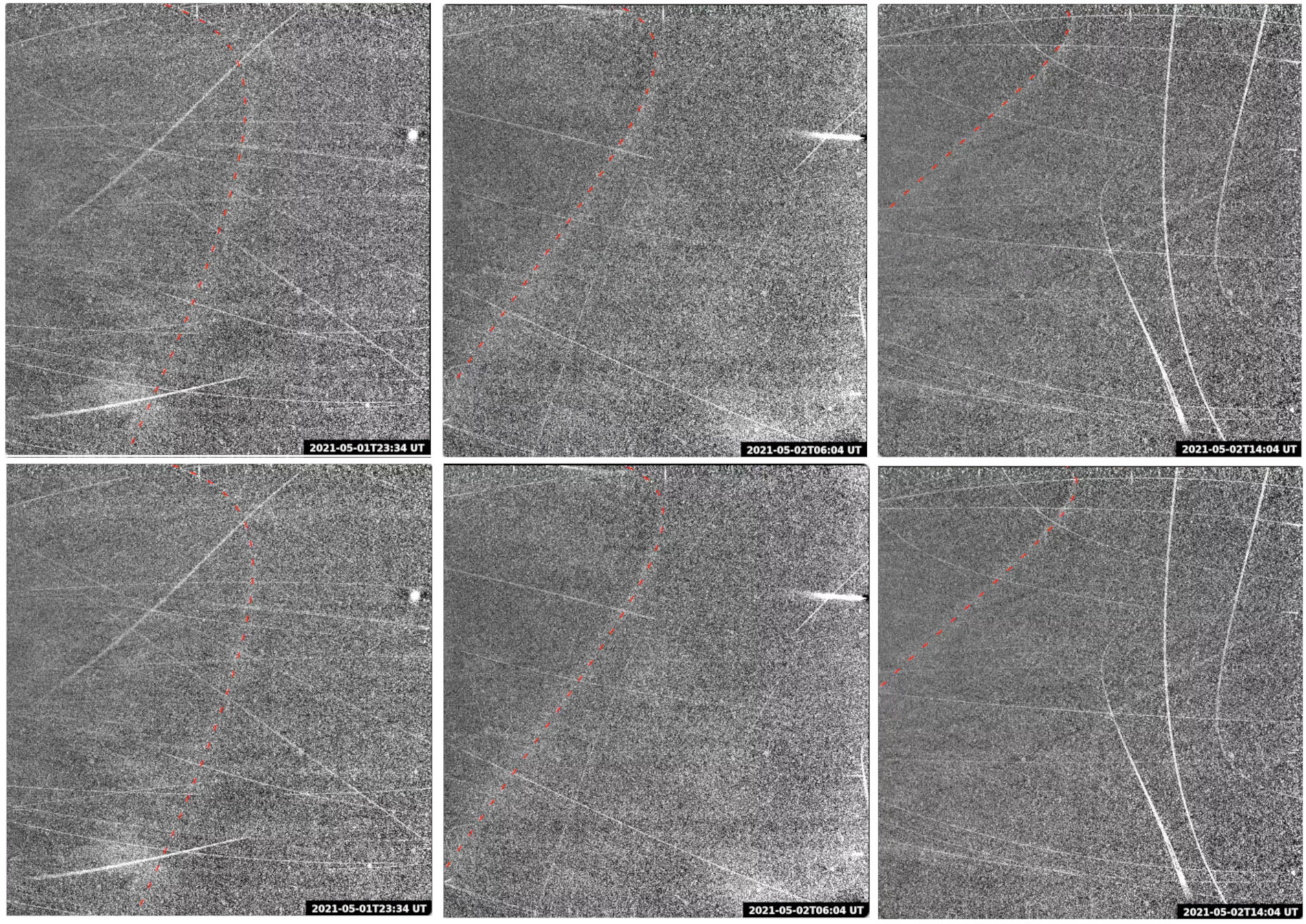}
\caption{Three WISPR-O LW-processed observations recorded on 2021 May 01 and 02 (E5), as indicated in the lower-right of each panel. The data in each column are identical. The top row shows the nominal trajectory of Phaethon (red dashed lines). The bottom row shows the nominal trajectory of Phaethon with the argument of periapsis reduced by exactly 1$^{\circ}$. The latter provides a far better visual correspondence to this portion of the dust trail.}
\label{fig:peri-adjust}
\end{figure}

Finally, we could consider that the trail was formed more recently than the Geminids. We have no knowledge of whether Phaethon has experienced any major outbursts over the estimated $\sim$2~kyr period since the Geminid formation, but we do know that it exhibits some kind of activity at perihelion \citep[e.g.][]{Jewitt2010}. We can say with near-certainty that the dust observed by WISPR has filled its orbit entirely, with no evidence for asymmetric brightness throughout the trail despite Phaethon itself having completed more than one orbit during the \textit{PSP} mission, and thus in this scenario the trail must be old. We could also arguably infer from Figure~\ref{fig:offset-plots} that some massively productive event(s) perhaps occur, or have occurred, with Phaethon at the $\sim$-10$^{\circ}$ True Anomaly location in its orbit, which would be just a few hours before perihelion. Ignoring any long-term evolution forces, we could naively assume the released dust to follow a very similar orbit to the parent but with a slightly smaller argument of periapsis corresponding to that earlier time of release. We investigated this idea by manually adjusting Phaethon's orbit by a few degrees in periapsis only, and visually comparing the resulting trajectory with the dust trail. Our best result is shown in Figure~\ref{fig:peri-adjust} which shows three WISPR-O observations (LW) on 2021 May 01 and 02 (E8), with the top row of panels showing a red dashed line on Phaethon's nominal orbit and the lower row of panels showing Phaethon's orbit with periapsis reduced by exactly 1$^{\circ}$. The latter appears almost perfectly centered on the observed dust trail orbit versus the obvious offset of Phaethon's trajectory. Further adjusting periapsis by 2$^{\circ}$ moved the plotted trajectory outside of the observed trail again. We also recreated the results of Figure~\ref{fig:offset-plots} and verified that the trend was much flatter, though still not perfectly flat, which unsurprisingly indicates that reducing periapsis by exactly 1$^{\circ}$ is not an ideal solution. There could be any number of adjustments to other orbital elements that may provide a better fit over the entire orbit, but do not have sufficient observations or any physical basis for suggesting alternative solutions at this time and thus will speculate no further. However, this result demonstrates that just minor adjustments are required to reconcile Phaethon's orbit with that of the dust trail - a result that may provide crucial clues to the origins and evolution of the dust we observe.

Thus, we are still left with many questions regarding the nature of the dust trail we are observing. Its relation to Phaethon seems beyond reasonable doubt, but the age, origin, and physical properties of its constituent dust remain entirely unclear. We hope future observations from several more encounters will enable us to more definitively model the dust and at least place tighter constraints on its age, mass, evolutionary pathway, and dust properties.

\subsection{Trail Photometry and Visibility}\label{sec:discuss-phot}

We have been able to conduct a much broader, multi-Encounter survey of photometric estimates of the trail than presented in Paper I. As with that study, avoiding dust streaks and bright stars in background apertures was uniformly challenging, but our results routinely returned a visual magnitude of V$\sim$16.1$\pm$0.3 per pixel, corresponding to a surface brightness of 26.1~mag~arcsec$^{-2}$. These results were supported by visual inspection of cross-trail profiles (\ref{app:A-photom}), which produced an alternative means by which we could estimate the brightness of the trail above the background noise in a given region of interest. This surface brightness value is somewhat fainter than that of Paper I, but is supported by better observations and methodology. The WISPR-I E1 observations analyzed in Paper I were incorporated into the present study and re-evaluated, finding them to be more consistent with those from the other orbits, albeit with larger uncertainties.

One of the interesting aspects of the photometry result is that the visual magnitude appears unrelated to the viewing parameters such as distance from the trail, trail width, true anomaly, or phase angle. If variation does exist as a function of these parameters, it is close to or within our error bars. We can draw a couple of conclusions from this. First, the dust must be quite uniformly distributed, which strongly supports our earlier conclusion that the dust has entirely filled its orbit. Second, the invariability of photometry as a function of true anomaly (and, by extension, its heliocentric distance) means that WISPR's detection of the dust trail is almost certainly simply the result of observing the scattering of sunlight from dust, versus some emission mechanism. For example, we briefly considered whether sodium emission in the dust could be a component of the detection, as the WISPR bandpass contains the sodium lines in which near-Sun comets are known to emit strongly \citep[e.g.][]{Knight2010,Battams2017}. However, this mechanism would require regular replenishment of sodium-rich dust, and it is challenging to conceptualize a reasonable mechanism by which ``fresh'' dust from Phaethon could migrate quickly into this separate trail orbit and still retain sufficient emitting (sodium) volatiles at true anomalies from $\sim$-20$^{\circ}$ -- 60$^{\circ}$. Thus we have some degree of confidence that sodium emission -- or indeed any emission (e.g. infrared, albeit not known to be in the WISPR bandpass) -- likely does not factor into the situation. Third, the apparent invariability to phase angle implies that the dust perhaps more like that of cometary grains than the asteroidal (`mini-Phaethon') assumption we nominally made in Paper I, as noted in Section~\ref{sec:mass}.

The lack of other visible cometary dust trails in the WISPR data hints that the Phaethon dust trail maybe be somewhat unique. In E1--E3, and very faintly in E3--E5, 2P/Encke's dust trail was observed at distances of $\sim$0.25 au -- 0.275 au from \textit{PSP}, but is no longer visible despite being within the field of view; we have not yet investigated the circumstances surrounding this, but increased distance from the trail appears to be the primary factor. The orbit of comet 96P/Machholz also passes through the fields of view during every encounter, with a closest approach of 0.0165 au occurring on 2020 September 25, and with \textit{PSP} now receding from 96P's orbit in all future encounters. Despite the 96P complex having no observed dust trail, the activity of the parent and frequent fragmentation of its children makes this system a reasonable analog for the Phaethon-Geminid system. While WISPR observed no discrete 96P-related dust structures during any of these periods, anomalous global brightening of the observations were noted in E6, E8, and E10, corresponding to the times at which \textit{PSP} was closest to 96P's nominal orbit \citep{Stenborg2022}. Further analysis of this phenomenon is underway in a separate study. On planetary scales, the WISPR instruments recently revealed evidence of a circumsolar dust ring in Venus' orbit \citep{Stenborg2021VenusRing}. We cannot comment on whether there is any connection between the properties of the Venus dust ring and the dust trails near Phaethon's orbit, although the very low plane angles of the spacecraft relative to Venus may again be a clue, with \textit{PSP} deviating less than 1$^{\circ}$ out of Venus' orbital plane for most of the mission duration.

There are certain counter-intuitive aspects of the trail's visibility in the WISPR data that warrant explanation. For example, in the L3 images (though not in the LW) the trail is always far less visible, if not entirely invisible, when approximately vertical in the WISPR-O observations. This is evident in the apparent gap in the trail offset/true anomaly plots of Figure~\ref{fig:offset-plots} at true anomaly values around 20-30$^{\circ}$ where we numerically struggle to capture it. This is a result of the computation of individual L2b background models, which involves the determination of the baseline brightness level at each image column. Vertical brightness features contribute in their entirety to the background level, while relatively thin, horizontal features do not. The excess signal loss does not meaningfully affect the photometric calibration by stars \citep{Hess2021}, but features already embedded in noise (e.g., the dust trail) may be affected.

We also found increased difficulty detecting the trail in WISPR-I, despite the WISPR-I crossing portions of the trail being close to \textit{PSP}, at seemingly innocuous phase angles, and in the camera that successfully observed the trail in E1-E3 under far less favorable conditions. This is likely a consequence of the dust trail density relative to the combined effect of local dust density of the Zodiacal cloud and the integration along the line of sight of the dust scattered light. In other words, the WISPR-I data are universally brighter due to their proximity to the Sun, and the trail in these data is effectively ``washed out'' as it becomes spread over, and embedded within, too many bright pixels. In the LW images there is also a relationship with how fast it moves through the field of view, with a faster apparent motion leading to clearer detections.

\subsection{Dust Trail Mass}\label{sec:discuss_dust_mass}

In Paper I we determined a mass estimate for the entire trail of $\sim(0.4-1.3){\times}10^{12}$~kg. In this study we have been able to revisit these mass calculations at a broad range of phase angles and spacecraft-trail distances, while also revisiting assumptions about both the mass (grain size) of the observed dust and its phase angle response.

The results presented in Table~\ref{tab:mass-calcs}, based upon 495 individual data points across seven of the nine encounters, present a consistent picture of the trail mass on the order of $10^{11}$~kg for 0.5~mm dust grains under the assumption of Phaethon-like dust grains, per Paper I. However, asteroidal phase functions can show significant responses across the $\sim$20$^{\circ}$-120$^{\circ}$ phase angle range observed by WISPR \citep{Muinonen2022, Ansdell2014} that should manifest clearly in photometry. We find no significant variation in trail brightness as a function of phase angle, implying a limitation in our 'mini-Phaethon' assumption. We therefore also assessed the same data set under the assumption of dust behaving more like cometary grains, which exhibit far less response in the range of phase angles observed by WISPR. This assumption leads to a factor of reduction in our mass estimates. According to \cite{Jewitt2010}, Phaethon's recent perihelion activity produces $\sim2.5{\times}10^{8}a_{mm}$~kg of material, where $a_{mm}$ is the dust radius in millimeters. As reported by \citep{Ye2018}, it is estimated that Phaethon would require approximately 250 years to fill an orbit with dust (as is the case with the dust trail in question), which for a 0.5~mm dust grain would represent a mass of $\sim10^{10}$. This is close to the WISPR trail mass assuming cometary grains, but does not address how or why thermally-fractured Phaethon dust would be cometary in nature.

Estimates for the total mass of the Geminid stream are of the order $10^{12}-10^{15}$~kg \citep{Ryabova2017}, which reconciles poorly with our mass estimates. However, Figure~\ref{fig:geminids-e7} shows clearly that the orbit of the dust trail in question bears only a slight resemblance to that of the Earth-crossing Geminids, and is unlikely a component of the Earth-encountering stream. Furthermore, our continued assumption from Paper I is that the bulk of the mass is contained in small grains -- an assumption that does not necessarily hold true for Geminids, as noted in Section 4 of \cite{Ye2018}. We could very likely construct a dust grain mass distribution that would force a much stronger overlap between the mass of the WISPR-observed trail and that of the Geminids. However, there currently seems little value in forcibly equating the mass of the WISPR-observed trail to that of the Geminids, when the former is clearly a separate structure from the latter. Furthermore, we have shown in Section~\ref{disc:morph_orb_prop} that the trail observed by WISPR lies not only extremely close to that of Phaethon, but also behaves more like a meteoroid stream released from its parent body shortly before the parent's perihelion. Based on this evidence it may be more appropriate to assume the WISPR dust trail to be related to perihelion activity of Phaethon, either ongoing or via a somewhat more recent large-scale outburst, than it is to be some remnant structure from the formation of the Geminids. Thus we do not feel that directly equating mass estimates of this near-Phaethon trail to those of the Geminids is an entirely valid equivalency.

The next several \textit{PSP} encounters will provide consistent viewpoints of the trail for over a year before the next orbital configuration change, providing a solid platform upon which to base a detailed dust modeling exercise and hopefully resolve more of these unanswered questions.

\section{Conclusions and Summary} \label{sec:concl}

We have presented a summary of white-light observations of a dust trail near the orbit of active asteroid 3200/Phaethon, recorded by the WISPR instrument on NASA's Parker Solar Probe during the first nine perihelion passages of the spacecraft. Following on from our initial publication announcing the discovery of the trail, we have presented updated analyses of the morphological and photometric properties of the trail and, crucially, are now able to provide some new insights into the orbital properties of the trail.

Specifically, we make the following key observations:
\begin{itemize}
    \item The observed dust trail does not share an orbit with its presumed parent, Phaethon, demonstrated by a clear visible separation between the dust trail and the orbit of Phaethon as a function of true anomaly. A naive modification to Phaethon's orbital parameters that reduces its argument of periapsis by exactly 1.0$^{\circ}$ returns a near perfect by-eye fit to the observed portion of the dust trail;
    \item We find the per pixel visual magnitude of the trail to be V$\sim$16.1$\pm$0.3 per pixel, corresponding to a surface brightness of 26.1~mag~arcsec$^{-2}$. These brightness values do not vary significantly as a function of true anomaly, indicating that the trail has an approximately uniform surface brightness and no detectable heliocentric dependence on brightness. This implies that the dust has entirely and uniformly filled its orbit, and that emission (e.g., sodium) does not play a significant role in the WISPR detection of the trail;
    \item We find an estimated mass of dust in the range $\sim10^{10}$~--~$10^{12}$~kg. This mass scales linearly with the assumed dust grain size, but still does not easily reconcile with the established Geminid stream mass without forcibly constructing an appropriate dust grain distribution, which does not seem valid given the significant orbital disparity between the WISPR trail and the Geminids.
    \item The visibility of the trail appears invariant to phase angle (within the ranges observed by WISPR), \textit{PSP} encounter date, encounter distance, and/or the heliocentric distance of the dust. This behavior is more consistent with the phase angle response of cometary dust grains. All observations of the trail have been recorded with the spacecraft within 2$^{\circ}$ of Phaethon's orbital plane, indicating that optical depth may be a key factor in WISPR's ability to detect the trail.
\end{itemize}

The lack of detection of this trail until the \textit{PSP} era is no mystery; in Paper I we demonstrated that the surface brightness of this trail presents an exceptionally challenging target for even the deepest existing sky surveys -- a conclusion which holds even more true in light of these new observations which have slightly reduced our estimate of the per-pixel visual magnitude and surface brightness. The trail appears sufficiently close to Phaethon's orbit, and certainly very dissimilar from the nominal Geminid orbit, that Earth would not pass through it and provide us with `atmospheric sampling' of the dust. The successful detection of the trail is a factor of \textit{PSP}'s fortuitous proximity, the large and highly sensitive WISPR pixels and, perhaps most crucially of all, the fact that WISPR observes from almost perfectly within the orbital plane of Phaethon.

The trail appears to be a distinctly separate feature from the Geminids, with orbital elements very similar to that of the parent. While this trail and the Geminids clearly share the same parent, they probably do not share the same formation circumstances and point-in-time of origin --- or at least were not formed during the same event. The data from E10 - E16 now brings the spacecraft to within 0.0137 au of Phaethon's orbit, and will provide valuable observations that would support a detailed dust-modeling effort to better characterize the true nature of this trail, and specifically the size and distribution of the dust grains, and its orbital properties.

The results presented here are of particular importance to the DESTINY+ mission \citep{Kruger2019} that will ultimately encounter Phaethon and likely also the dust structure observed in the WISPR observations. These results, and similar WISPR-based publications \citep[e.g.,][]{Stenborg2021VenusRing, Stenborg2021DDZ} regarding dust observations, also highlight the unique value of having sensitive wide-field white-light imaging instruments operating inside the inner solar system.

\begin{acknowledgments}
\centering ACKNOWLEDGEMENTS

We would like to thank the anonymous referee for their helpful comments with this manuscript. Parker Solar Probe was designed, built, and is now operated by the Johns Hopkins Applied Physics Laboratory as part of NASA's Living with a Star (LWS) program (contract NNN06AA01C). Support from the LWS management and technical team has played a critical role in the success of the Parker Solar Probe mission. The Wide-Field Imager for Parker Solar Probe (WISPR) instrument was designed, built, and is now operated by the US Naval Research Laboratory in collaboration with Johns Hopkins University/Applied Physics Laboratory, California Institute of Technology/Jet Propulsion Laboratory, University of Gottingen, Germany, Centre Spatiale de Liege, Belgium and University of Toulouse/Research Institute in Astrophysics and Planetology. G.S. was supported by WISPR Phase-E funds. KB, AL, BG, ML and RH were supported by the NASA Parker Solar Probe WISPR Project.

\end{acknowledgments}

\software{astropy \citep{Astropy13}
          }

\newpage
\appendix

\section{WISPR Image Processing}\label{app:A1}

\begin{figure}[ht!]
\centering
\includegraphics[width=0.9\textwidth]{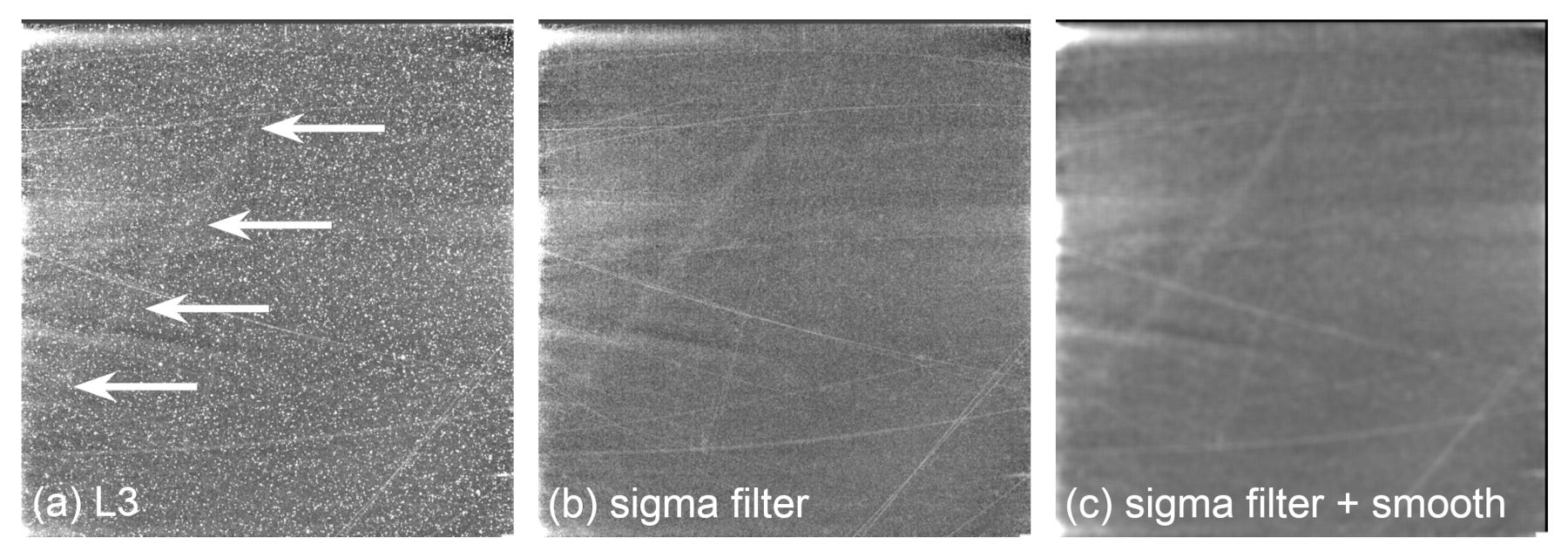}
\caption{A WISPR-O observation recorded on 2021 May 02 07:34UT processed in three different formats. Figure~\ref{fig:processing}(a) shows the L3 version of this image -- that is, a fully calibrated image with background image removed. Figure~\ref{fig:processing}(b) shows the L3 image with a sigma filter applied. Figure~\ref{fig:processing}(c) shows the L3 image with a sigma filter and smooth filter applied. The dust trail can be seen crossing diagonally from the top-center towards the lower-left corner, more clearly in panels (b) and (c).}
\label{fig:processing}
\end{figure}

Depending on the nature of the analysis in question, different image processing techniques were required to enhance the visibility of the dust trail in the WISPR observations. For illustration, Figure~\ref{fig:processing} shows a background-removed snapshot of a WISPR-O observation taken on 2021 May 02 07:34UT, processed and enhanced in three different ways. Namely, Figure~\ref{fig:processing}(a) shows the corresponding L3 image, i.e., a fully L2 calibrated image with the background scene (primarily zodiacal dust) removed as detailed in \cite{Battams2020} and properly scaled in brightness to highlight the visibility of the trail, which is indicated with white arrows. Figure~\ref{fig:processing}(b) shows the L3 image with a ``sigma filter'' applied. The sigma filter is well suited to minimize the effect of bright point-like sources, and hence is particularly useful to minimize the effect of the star field and saturating planets. Finally, Figure~\ref{fig:processing}(c) shows the sigma-filtered L3 image with a low-pass filter applied to smooth the texture of the snapshot. Not shown here, but shown in later Figures, is an alternate processing technique we call ``LW'' processing, which was developed to improve the visibility of features that move across the FOV of the WISPR telescopes.

Unlike the background determination approach developed to create the L3 data sets, where the background scene is determined out of each individual image, this alternative approach exploits the time domain to create a model of the median brightness at each pixel location at each time instance. The median brightness level is computed in the time domain with a boxcar window equivalent to about nine time instances. Therefore, long-lived, stationary structures in the FOV will be removed after background subtraction (e.g., the dominant F-corona background scene and pseudo-stationary K-corona structures as streamers), hence enhancing features that appear as non-stationary in the FOV of the WISPR telescopes. While there is a concern that slower-moving structures could be unintentionally removed, the dust trail that we study here moved rapidly across the field of view, particularly during more recent encounters, and thus these effects should be quite minimal.

\section{Data stacking}\label{app:A}

As noted in Section~\ref{sec:analysis}, the noise-level signal from the dust trail coupled with the frequent dust streaks in the WISPR data mean that photometric and morphological analyses of the trail are largely impossible to routinely perform on individual images. Instead we must apply both image processing and averaging techniques to improve the signal-to-noise ratio to an acceptable level such that the trail is photometrically distinct from the background noise. As noted in Appendix~\ref{app:A1}, the image processing applied to the data was different based on the nature of the analysis. Regardless of data pre-processing, however, the process for data stacking remained the same, and is summarized as follows.

\subsection{Region of Interest (ROI) Selection}

First, we identify a sequence of at least five consecutive WISPR images in which the dust trail is not heavily impacted by dust streaks. Next, based on the pixel positions that would be directly on Phaethon’s orbit, a section of the trail (region of interest, or ROI) is identified by-eye, with the selection criteria again being that the target portion of the trail is not too heavily polluted by dust streaks or bright planets during that five (or more) image sequence. The width (cross-trail) of this ROI is selected to be approximately 4.2 degrees (approximately 75 pixels wide), centered on Phaethon's orbit, with this width being sufficient in all encounters to encompass the breadth of the trail. The length of the ROI (along the trail) is selected such that the extreme bounds are based on the simulated rates of Phaethon's motion over the course of nine hours as seen by WISPR. That is, if Phaethon itself were in that field of view, it would exactly transit the ROI in nine hours.  Examples of such regions of interest are shown as the white boxes in Figures~\ref{fig:trail-profile-smoothed-fit} and \ref{fig:trail-profile-raw}.

The selection of a time interval of nine hours (i.e., Phaethon's motion) was largely arbitrary. In practice, the choice of the ROI length does not have an appreciable impact on the remainder of the process as long as the ROI is not excessively small or long. We simply found that nine hours of  Phaethon's motion returned ROI's of a reasonable length. Alternative criteria could be an apparent length in degrees, a length based on some angular range in true anomaly, or even some fixed length in pixels. These would work equally as well, but with the same shortcomings -- namely that the ROI's can never capture the changing viewing geometry of the trail in a systematically fair way. For example, if the ROI size was based on a fixed pixel size, ROI's placed at large true anomaly locations would encompass more of the trail than ROI's at small anomaly locations.

Thus, based on tests of several alternatives, we opted for ROI's based on Phaethon's simulated motion simply because it was the metric that worked conveniently with our algorithms. The noise in the data, and uncertainties in Gaussian-fitting, were overwhelmingly more dominant factors than ROI sizes.

\subsection{Cross-trail Profiles}

\begin{figure}[ht!]
\centering
\includegraphics[width=0.35\textwidth]{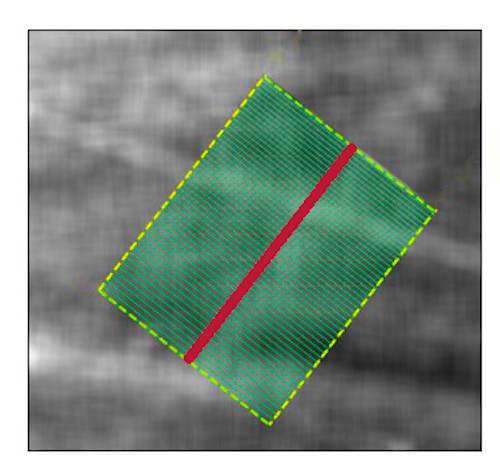}
\caption{Illustration of defined region of interest (ROI; yellow dashed boundary) with the orbit of Phaethon (solid red line) passing through the center of the ROI, with the dust trail very faintly visible following this line. Green lines across the ROI illustrate the series of cross-trail profiles that we define and use to extract signal from the ROI. Data from these lines are averaged (median) to produce a single cross-trail profile for that ROI. Note that this image is just illustrative of the process; actual ROI size and number of cross-trail lines will differ, as noted in the manuscript.}
\label{fig:roi-lines}
\end{figure}

The next step was to construct a series of evenly-spaced lines along the ROI, perpendicular to the orbit of Phaethon. These perpendicular lines are centered on Phaethon’s orbit and extend out on each side past the region where the trail can be visually detected in the images (approximately 2.2$^\circ$). This is illustrated in Figure~\ref{fig:roi-lines}, which shows a ROI (yellow outline) centered along the orbit of Phaethon (red line), with a series of perpendicular (green) lines defined perpendicular to Phaethon's orbit. Along each of these perpendicular lines we then identify a series of evenly-spaced sample points from which the actual data values are extracted.

This method required that the perpendicular lines were evenly spaced, and likewise that the sample points along each line were also equally spaced. We evaluated two different approaches for calculating this spacing. The first and simplest approach was to create evenly spaced points based on the two-dimensional geometry of the flat image, (i.e. the x,y pixel separation). The alternate approach was to take into consideration the curvature of the field of view as seen from the WISPR camera and first convert x,y pixel locations to physical sky coordinates, and then space them accordingly in angular units before mapping back to x,y pixel coordinates again. Again, both methods were found to return effectively the same results, and thus we chose to adopt the simple flat geometry (x,y pixel) method, which was computationally faster and simpler.

After constructing the perpendicular lines and sample points, intensity values are extracted at each sample point along each line, resulting (computationally) in a series of one-dimensional arrays representing the perpendicular lines. These arrays can then be averaged (median, to smooth out the contributions from outliers such as stars) and plotted to produce a final cross-trail profile for that ROI, such as that shown in panels (b) and (c) of Figures~\ref{fig:trail-profile-smoothed-fit}, \ref{fig:trail-profile-smoothed-fit_LW}, and \ref{fig:trail-profile-raw}.

\subsection{Data Pre-processing}

As noted in the manuscript (Section~\ref{sec:wispr-obs}), the nature of the analysis in question --specifically, quantitative versus qualitative, influenced the choice of the pre-processing approach. Qualitative analyses that simply required the estimation of the Gaussian model parameters (e.g., for statistical purposes) may benefit from the ''LW'' processing (Figure~\ref{fig:processing}(c)) or very low-noise `sigma plus smooth' filtering ((Figure~\ref{fig:processing}(b)). These resulted in smoother profiles and more consistent Gaussian fit results. Also, both techniques generally returned the same Gaussian parameters (to within some degree of tolerance) when operating on the same data, but we did note that certain periods of observation produced clearer profiles with one method versus the other. Generally speaking, the LW algorithm was preferred for observing the trail at larger true anomalies, whereas the sigma plus smoothing algorithm was preferred at smaller true anomalies. The reason for this is not entirely clear, but most likely related to the different ways in which the two techniques handle the background (F-coronal) noise removal. But again, results between the two were largely consistent.

Quantitative analysis, i.e. that requiring photometry be preserved, required the much noisier L3 data product, but the sigma filter plus smoothing could still be employed here if pixel-level photometry was not required. For pixel-level photometry - i.e., determining the trail brightness in small areas, just the L3 data alone was used, resulting in far less consistent Gaussian fits to that data. Additionally, we investigated using Gaussian plus constant term, and Gaussian plus linear fits instead of a simple Gaussian, in order to provide a more consistent estimate of the amplitude (and thus the photometric excess from the dust). However, we found that the background noise was both so strong and so variable that there was little consistency in the fit parameters of these extra terms. That is, even in consecutive datasets, the background noise floor could be a strong positive gradient in one profile plot, and a strong negative gradient in another. This was dictated entirely by the background scene, whether it be solar streamers, CME's, bright stars, dust streaks, or just the global radial brightness gradient inherent to these observations. Therefore, this L3-only technique was only employed as a secondary method to support our aperture-photometry trail brightness estimates (Section~\ref{sec:discuss}), and only on a specific case-by-case basis.

\subsection{Cross-trail profile examples}

\begin{figure}[ht!]
\centering
\includegraphics[width=0.9\textwidth]{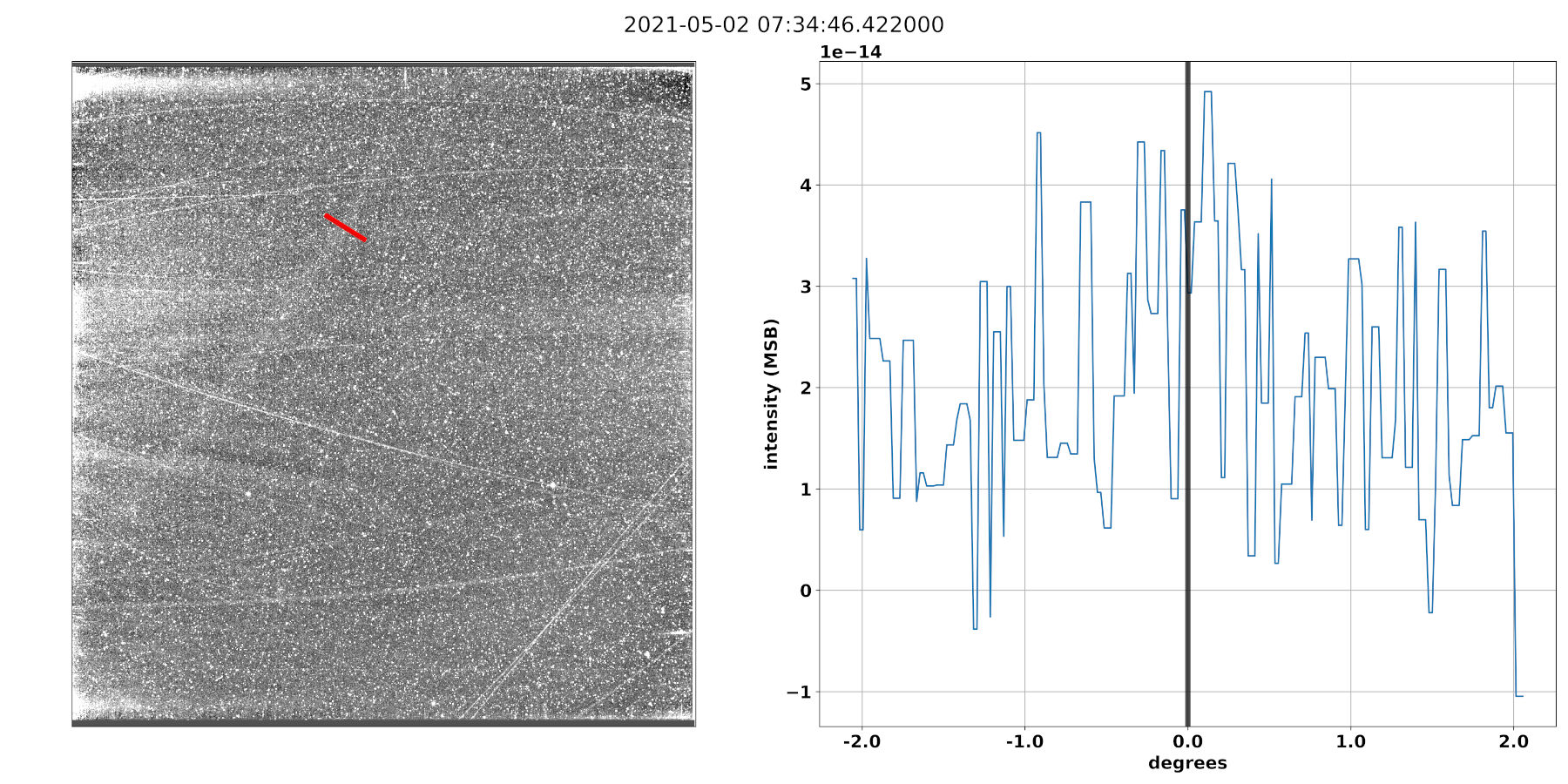}

\caption{A single cross-trail line profile (right panel) extracted along a $\sim$4$^{\circ}$ line ($\sim$71 pixels long, red line in left panel), bisecting the dust trail and Phaethon's orbit, as seen in WISPR-O (left panel) on 2021 May 02 07:34UT. The profile is centered on the orbit of Phaethon, as indicated by the dark vertical line at 0$^{\circ}$. The data intensity is in calibrated mean solar brightness units, and the width of the red line has been exaggerated to aid visibility. Contrast in the image has been manually changed to enhance the visibility of the trail in print copies of this manuscript.}
\label{fig:single-profile}
\end{figure}

The following Figures provide examples of the result of extracting cross-trail profiles from the WISPR observations using various stacking and pre-processing techniques.  To illustrate this fact, we show in Figure~\ref{fig:single-profile} a profile extracted from a $\sim$4$^{\circ}$ line ($\sim$71 pixels long, red line in left panel), bisecting the dust trail and Phaethon's orbit, as seen in an L3 WISPR-O (left panel) image on 2021 May 02 07:34UT. The dark vertical line at 0$^{\circ}$ indicates the location of Phaethon's orbit, and the y-axis intensity is calibrated in units of mean solar brightness (MSB, or $B_\sun$). The red line is exaggerated in size, and the contrast of the image manually changed, to aid visibility in print copies of this manuscript. We also note that the few spurious negative pixel values are a consequence of overcompensation from the background models used in production of the L3 data; they do not affect any of the analyses we presented here. As we will demonstrate in later Figures, the signal from the dust trail is on the order of 0.5$\times$10$^{-14}$ $B_\sun$, which is clearly insufficient to be prominent in single-profile observations such as this, and even smoothing and/or compressing individual images does not make it possible to extract meaningful cross-trail profiles. Thus, to increase the SNR, we implemented a stacking approach (see Appendix~\ref{app:A}), which enables us to examine and characterize the cross-section profile of the dust trail as it progresses across the FOV of WISPR-O.

Briefly, the technique relies on (1) defining a region of interest (ROI) centered along Phaethon's orbit, (2) extracting multiple adjacent cross-trail profiles within the ROI, and (3) stacking (averaging) these cross-trail brightness profiles to improve the SNR of the cross-sectional brightness of the trail. An empirical model of the resulting profile is obtained by fitting a Gaussian to the so-obtained measurements. This model allows us to measure the full-width at half-maximum (FWHM), the amplitude, and the location of the central peak, which in turn enables the estimation of the apparent offset of the trail from the projection of Phaethon's orbit in WISPR-O images. Since the distance between \textit{PSP} and Phaethon's orbit is known, both the angular plane-of-sky distance and the true apparent physical distance can be then determined. This technique also offers an alternate means for estimating photometry, as discussed in Section~\ref{sec:photom}.

\begin{figure}[ht!]
\centering
\includegraphics[width=0.9\textwidth]{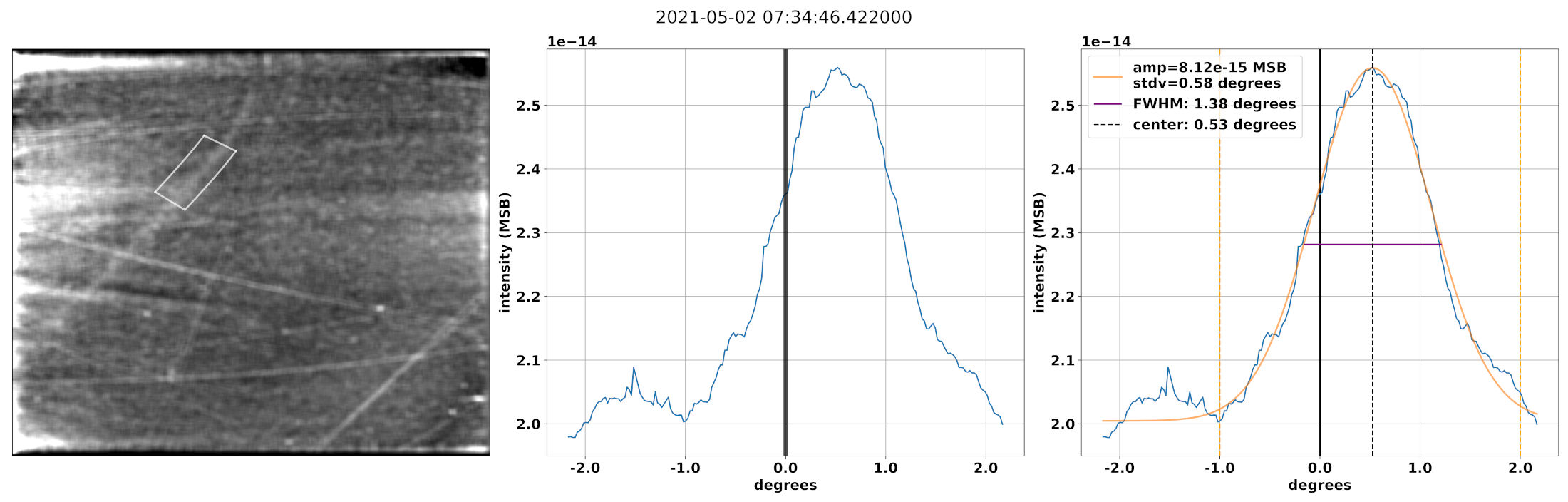}
\caption{Illustration of a cross-trail profile extracted from an L3 WISPR-O image, recorded on 2020 June 07 (first panel), and  processed with the sigma filter plus smoothing algorithm. The white box outlines the region of interest in which 396 cross-trail profiles were extracted. The second shows the resulting profile, averaged from these 396 profiles, and the third panel shows the result of a Gaussian fit to the profile, and the resulting fit parameters. The intensity is recorded in physical mean solar brightness units, but note that the smoothing filter can be detrimental to pixel-level photometry.}
\label{fig:trail-profile-smoothed-fit}
\end{figure}

In practice this technique works well when the images are mostly free of streaks in the ROI. The presence of point-like sources on or nearby the trail can be minimized by the use of a sigma-filter prior to stacking. A low-pass (smoothing) filter can also be applied to further reduce noise and hence improve the visibility of the trail and return much smoother cross-trail profiles (see, e.g., Figure~\ref{fig:processing}). This is illustrated by Figure~\ref{fig:trail-profile-smoothed-fit}, which shows a WISPR-O L3 image recorded on 2020 June 07 (E5), processed with both the sigma filter and a smoothing filter. (The contrast has also been stretched here to better show the trail in print copies of this manuscript.) The white box in the first panel outlines the region in which 396 cross-trail profiles were extracted, with the Phaethon dust trail seen quite clearly crossing through the ROI. The center panel shows the resulting averaged profile, with y-axis representing the averaged intensity in the ROI. We will note here that the photometry of the data is globally still preserved by this smoothing algorithm, but pixel-level photometry is not, and thus caution must be used when employing such a technique with these observations. The x-axis represents angular distance in degrees from Phaethon's orbit, which is indicated by the thick vertical line at 0$^{\circ}$. The third panel shows the resulting Gaussian fit to this profile, including the FWHM, amplitude, and central peak location. In this example, the dust trail appears offset by approximately $\sim$0.53$^{\circ}$ from Phaethon's orbit according to the Gaussian parameters.

\begin{figure}[ht!]
\centering
\includegraphics[width=0.9\textwidth]{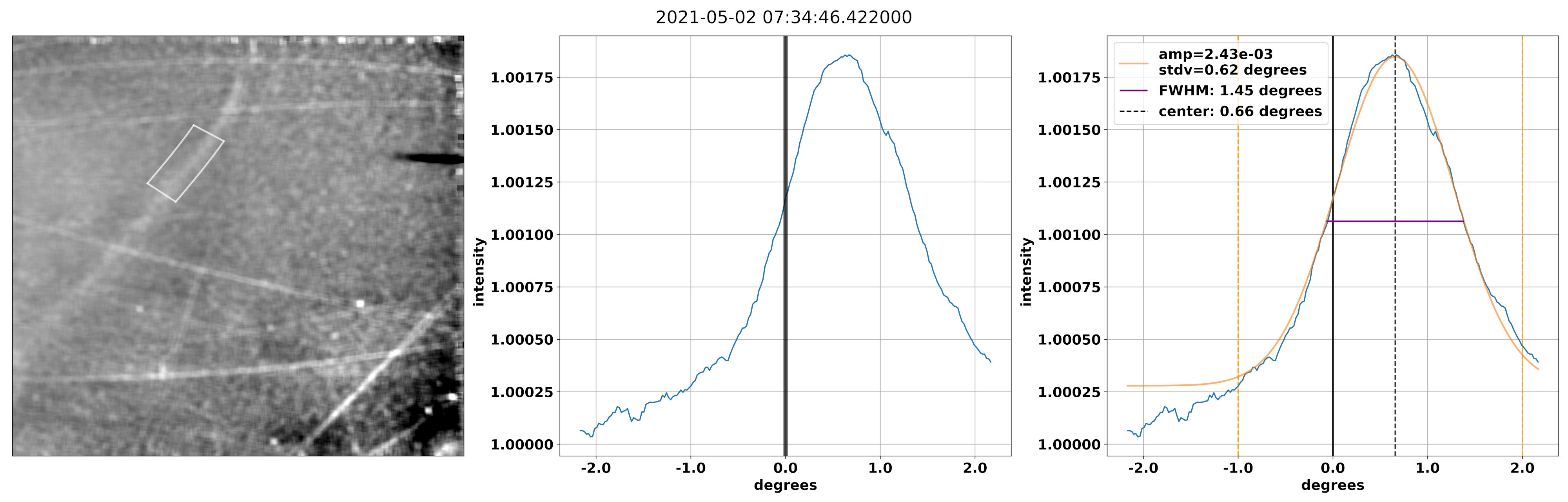}
\caption{This illustration is the same as that shown in Figure~\ref{fig:trail-profile-smoothed-fit}, except the profile was extracted from an LW WISPR-O image. The sigma filter plus smoothing algorithm was applied again. The intensity is scaled to a minimum value of zero as a result of employing the LW algorithm, thus quantitative interpretation of intensity is not possible.}
\label{fig:trail-profile-smoothed-fit_LW}
\end{figure}

In some cases we found the trail visibility was enhanced by use of the LW filter in addition to the sigma and smoothing filters. This is illustrated in Figure~\ref{fig:trail-profile-smoothed-fit_LW}, where we used the LW-processed version of the WISPR observation (instead of the L3 version) with the same ROI defined in Figure~\ref{fig:trail-profile-smoothed-fit}. (Again, the contrast on this image has been stretched to improve its visibility in print.) This process often produced a slightly smoother profiles but the inclusion of the LW algorithm completely destroys photometry and thus again must be employed with caution. As seen here, the two algorithms produced slightly different values for the trail offset, with this latter algorithm returning $\sim$0.66$^{\circ}$. No systematic bias was observed in offset values returned by these algorithms, with at most one or two tenths of a degree difference between their returned offset in almost all circumstances. Thus, it was deemed acceptable to use these algorithms interchangeably in this study, though we may return to this in the future to investigate why certain processing was preferable at different points in each Encounter. We describe later in the manuscript the circumstances under which the LW-based version of the processing was used in preference to the L3.

\subsection{Obtaining Photometric Estimates from Cross-Trail Profiles}\label{app:A-photom}

\begin{figure}[ht!]
\centering
\includegraphics[width=0.9\textwidth]{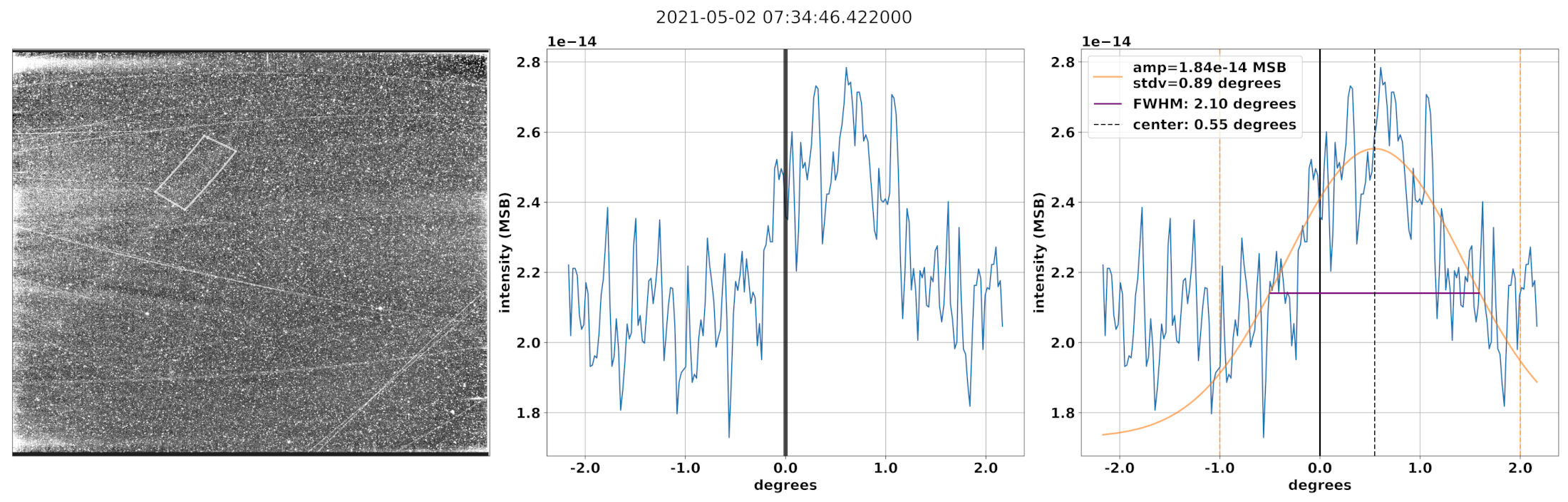}

\caption{Illustration of a photometry-preserving averaging method for estimating the visual magnitude of the dust trail. The first panel shows an L3 image recorded on 2021 May 02 (E5) marked with a region of interest (ROI) centered along Phaethon's orbit. The white box outlines the region of interest in which 396 cross-trail profiles were extracted. The second panel shows the resulting profile, averaged from these 396 profiles, and the third panel shows the result of a Gaussian fit to the profile, and the resulting fit parameters. The intensity is recorded in physical, calibrated units and thus can be directly interpreted from such plots.}
\label{fig:trail-profile-raw}
\end{figure}

Figure~\ref{fig:trail-profile-raw} illustrates how cross-trail profiles can be used to estimate the photometry of the dust trail. The first panel shows an L3 WISPR-O image recorded on 2021 May 02 (E8) centered along Phaethon's orbit, with the white box outlining the region of interest in which 396 cross-trail profiles were extracted. The second panel shows the resulting profile, averaged from these 396 profiles, and the third panel shows the result of a Gaussian fit to the profile, and the resulting fit parameters. As we are using just the L3 data product, the intensity is recorded in physical, calibrated units. In this particular example, the profile returns an offset of $\sim$0.55$^{\circ}$, which is consistent with those returned by the smoothing-based processing of this image shown in Figures~\ref{fig:trail-profile-smoothed-fit} (0.53$^{\circ}$) and \ref{fig:trail-profile-smoothed-fit_LW} (0.66$^{\circ}$). That said, this is a fortunate choice of example, as in many cases, fitting Gaussians to these raw profiles returned widely varying results, including numerous spurious values.

For photometric purposes, we can assume the total signal in the third panel of Figure~\ref{fig:trail-profile-raw} to be a sum of \textit{background noise + dust signal}, then by-eye we can approximate the dust signal to be the peak of the central portion of the profile and thus the background noise as being the continuum outside of that range. In the example provided, we would estimate the background to be 2.1$\times$10$^{-14}$ $B_\sun$, and the peak of the dust trail to be 2.6$\times$10$^{-14}$ $B_\sun$. Thus via simple subtraction we can see in this example that the dust trail peaks at $\sim$0.5$\times$10$^{-14}$ $B_\sun$ above the background continuum. This value corresponds to a visual magnitude of 16.2, which is entirely consistent with the photometry obtained by the background subtraction technique. This exercise could also have been accomplished via fitting with a Gaussian plus constant term, to provide a better representation of the Gaussian amplitude. However this was an idealized example and, as we have noted, Gaussian fits with either constants or linear terms often produced inconsistent results due to the highly variable noise floor.

\bibliography{references}{}
\bibliographystyle{aasjournal}

\end{document}